\documentclass[twocolumn,prb,showpacs,preprintnumbers,amsmath,amssymb,superscriptaddress]{revtex4-1}
\usepackage{graphicx}
\usepackage{setspace}

\usepackage{color}
\usepackage{ulem} 
\graphicspath{{./figures/}}

\usepackage{amsmath,amsthm,amssymb}
\usepackage{mathrsfs}

\usepackage[%dvipdfmx,
colorlinks=true, citecolor=blue, urlcolor=blue, linkcolor=blue,
setpagesize=false, bookmarks=false, breaklinks=true
]{hyperref}

\newcommand{\hc}{\hat{c}}
\newcommand{\hd}{\hat{d}}

\newcommand{\hh}{\hat{h}}
\newcommand{\hb}{\hat{b}}
\newcommand{\heta}{\hat{\eta}}
\newcommand{\hs}{\hat{s}}
\newcommand{\hH}{\hat{H}}
\newcommand{\hA}{\hat{A}}
\newcommand{\hB}{\hat{B}}
\newcommand{\hn}{\hat{n}}

\newcommand{\hS}{\hat{S}}

\newcommand{\hM}{\hat{M}}

\newcommand{\eqq}[1]{\begin{align} #1 \end{align}}

\begin{document}
\title{High-harmonic generation in one-dimensional Mott insulators}

\author{Yuta Murakami}
\affiliation{Department of Physics, Tokyo Institute of Technology, Meguro, Tokyo 152-8551, Japan}
\author{Shintaro Takayoshi}
\affiliation{Department of Physics, Konan University, Kobe 658-8501, Japan}
\author{Akihisa Koga}
\affiliation{Department of Physics, Tokyo Institute of Technology, Meguro, Tokyo 152-8551, Japan}
\author{Philipp Werner}
\affiliation{Department of Physics, University of Fribourg, 1700 Fribourg, Switzerland}
\date{\today}

\begin{abstract}
We study high-harmonic generation (HHG) in the one-dimensional Hubbard model in order to understand its relation to elementary excitations as well as the similarities and differences to semiconductors. The simulations are based on the infinite time-evolving block decimation (iTEBD) method and exact diagonalization.
We clarify that the HHG originates from the doublon-holon recombination, and the scaling of the cutoff frequency is consistent with a linear dependence on the external field.
We demonstrate that the subcycle features of the HHG can be reasonably described
by a phenomenological three step model for a doublon-holon pair.
We argue that the HHG in the one-dimensional Mott insulator is closely related to the dispersion of the doublon-holon pair with respect to its relative momentum, which is not necessarily captured by the
single-particle  spectrum due to the many-body nature of the elementary excitations. 
For the comparison to semiconductors,
we introduce effective models obtained from the Schrieffer-Wolff transformation, i.e. a strong-coupling expansion, which allows us to disentangle the different processes 
involved in the Hubbard model:
intraband dynamics of doublons and holons, interband dipole excitations, and spin exchanges.
These demonstrate the formal similarity of the Mott system to the 
semiconductor models in the dipole gauge, 
and reveal that the spin dynamics, which does not directly affect the charge dynamics, can reduce the HHG intensity.
We also show that the long-range component of the intraband dipole moment has a substantial effect on the HHG intensity, 
while the correlated hopping terms for the doublons and holons 
essentially determine the shape of the HHG spectrum.
A new numerical method to evaluate single-particle spectra within the iTEBD method is also introduced.
\end{abstract}

\maketitle

\section{Introduction}

High-harmonic generation (HHG) is an intriguing nonlinear phenomenon
originating from strong light-matter coupling.
It has been originally observed and studied in atomic and molecular gases,\cite{Corkum1993PRL,Lewenstein1994} and is used in attosecond laser sources as well as 
spectroscopies.\cite{Krausz2009RMP} 
Recently, HHG has been observed in semiconductors and semimetals,\cite{Ghimire2011NatPhys,Schubert2014,Luu2015,Vampa2015Nature,Langer2016Nature,Hohenleutner2015Nature,Ndabashimiye2016,Liu2017,You2016,Yoshikawa2017Science,Hafez2018,Kaneshima2018,Yoshikawa2019,Matsunaga2020PRL} 
which expands the scope of this field to condensed matter systems.\cite{Golde2008,Higuchi2014,Vampa2014PRL,Vampa2015PRB,Wu2015,Otobe2016,Luu2016,Ikemachi2017,Hansen2017,Tancogne-Dejean2017b,Tancogne-Dejean2017,Osika2017,Chacon2018arXiv,Tamaya2016,Floss2018,Markus2020SOC,Kemper2013b,Almalki2018PRB,Yu2019PRA,Yu2019PRB2,Orlando2018,Orlando2019,Chinzei2020,Ikemachi2018}
It has been clarified that
the HHG in semiconductors and semimetals can be well understood from the dynamics of 
independent electrons in the periodic lattice potential (single-particle picture),
since the interactions among the charges can be neglected to a first approximation.
\cite{Golde2008,Higuchi2014,Vampa2014PRL,Vampa2015PRB,Wu2015,Otobe2016,Luu2016,Ikemachi2017,Hansen2017,Tancogne-Dejean2017b,Tancogne-Dejean2017,Osika2017,Chacon2018arXiv,Tamaya2016,Floss2018,Markus2020SOC}
Hence HHG can be a powerful tool to detect the dispersion of the 
conduction bands\cite{Ghimire2011NatPhys,Luu2015,Li2020HHG} as well as 
the Berry curvatures\cite{Luu2018Amorphas} (HHG 
spectroscopy of condensed matter systems).
Further exploration of the HHG has been carried out both experimentally and
theoretically in various other systems such as amorphous 
materials,\cite{You2017Natcom,Luu2018Amorphas} liquids, 
\cite{Heissler2014NJP,Luu2018Liquid}, strongly correlated 
systems,\cite{Silva2018NatPhoton,Murakami2018PRL,Murakami2018PRB,Markus2020,Tancogne-Dejean2018,Ishihara2020,vaskivskyi2020} and spin or multiferroic systems.\cite{Takayoshi2019PRB,Ikeda2019PRB}

In this paper, we focus on strongly correlated electron systems (SCES), which provide a potentially interesting playground for HHG.
For example, it is known that the third-harmonic generation (THG) signal 
is relatively large in one-dimensional (1d) Mott insulators.\cite{Kishida2000,Kishida2001,Mizuno2000}
Furthermore, the HHG spectrum sensitively reflects the properties of the 
phase and can be utilized as a detector of phase transitions in SCESs,
{\it e.g.}, the photo-induced melting of Mott insulators.\cite{Silva2018NatPhoton}
One of the significant characteristics in SCESs is the existence of many-body elementary excitations,
which are distinct from conventional electron- and hole-excitations in semiconductors.
The dynamics of such elementary excitations under strong fields can result in nontrivial HHG in SCESs.
For HHG in Mott insulators, the signature of doublon and holon dynamics \cite{Murakami2018PRL,Murakami2018PRB} and string states characteristic of SCESs \cite{Markus2020} have been discussed.
Moreover, in dimer Mott systems, the dynamics of the kinks and anti-kinks has been pointed out to be the origin of HHG.\cite{Ishihara2020} 
However, the understanding of HHG in SCESs is still limited
compared to semiconductors or semimetals, and fundamental questions 
remain to be answered: i) How is HHG connected to the dynamics of 
the elementary excitations in SCESs and what information can be obtained 
from it, ii) How is the HHG in SCESs similar to or different from that of semiconductor systems, and iii) What is the role of the characteristic degrees of freedom in SCESs such as spins?

To address these fundamental questions, we study the 1d Mott insulator described
by the single-band Hubbard model at half filling.
The calculations are based on the infinite time-evolving block 
decimation (iTEBD)\cite{Vidal2007PRL} and the exact diagonalization (ED) methods.
We demonstrate that the doublon-holon recombination is dominant for the HHG and
that the cutoff frequency scaling is consistent with a linear scaling against the field strength. 
In addition, we present the subcycle features of the HHG spectrum in the 
Mott insulator. We show that it can be reasonably described in 
terms of a semiclassical trajectory analysis for a doublon-holon pair, 
 whose kinetics is ruled by the dispersion of the doublon-holon pair with respect to its relative momentum.
This dispersion is obtained from the Bethe ansatz results, and can be reasonabely extracted from the single-particle spectrum in the case of the Hubbard model.
However, we point out that in general the dispersion of the doublon-holon pair is not necessarily captured in the single-particle spectrum due to the many-body nature of the elementary excitations.

The formal similarities and differences compared to the 
semiconductor model and the various processes 
involved in the HHG of the Hubbard model become clear in the effective models derived from the time-dependent Schrieffer-Wolff transformation.\cite{MacDonald1988}
Using these models, we show that the spin dynamics, which has no analogue in semiconductor systems, reduces the HHG intensity.
We also reveal the importance of the long-range component of the ``dipole moment'' between doublon and holon bands for the HHG intensity, as well as the role of the correlated hopping of doublons and holons for the shape of the HHG spectrum.  

This paper is organized as follows.
In Sec.~\ref{sec:2}, we introduce the Hubbard model.
Then, we briefly explain a new method to evaluate the single-particle spectrum in iTEBD
and derive the effective models. 
The numerical results are presented in Sec.~\ref{sec:results}.
We discuss the single-particle spectrum, which is used for the semiclassical trajectory analysis,
and present the HHG spectrum obtained from iTEBD.
The analysis based on the effective model is also shown. 
Finally, we conclude the discussions in Sec.~\ref{sec:conclude}.

\section{Formulation}\label{sec:2}
\subsection{Model and Method}\label{sec:model}
We consider the 1d Hubbard model, with Hamiltonian
\begin{align}
\hH(t) &= -v\sum_{i,\sigma} [e^{-iA(t)} \hc^\dagger_{i,\sigma} \hc_{i+1,\sigma} + h.c.]  -\mu \sum_i \hn_i\nonumber \\
&+ U\sum_i  \hn_{i,\uparrow}\hn_{i,\downarrow}  +h_z \sum_i (-)^i \hs_{z,i}, \label{eq:Hubbard}
\end{align}
where $\hc^\dagger_{i,\sigma}$ is a creation operator of an electron with spin $\sigma$ at site $i$,
$\hn_{i,\sigma} = \hc^\dagger_{i,\sigma} \hc_{i,\sigma},\; \hn_i = \hn_{i,\uparrow} + \hn_{i,\downarrow}$,
and $\hs_{z,i}=\frac{1}{2}( \hn_{i,\uparrow} - \hn_{i,\downarrow})$.
$v$ is the hopping parameter and $U$ is the onsite Coulomb interaction.
$\mu$ is the chemical potential, which is set to $U/2$, i.e., half filling. 
The last term in Eq.~(\ref{eq:Hubbard}) represents 
the effect of a staggered magnetic field $h_z$.
In the following simulations with iTEBD, we apply a small staggered 
magnetic field $h_z=10^{-3}v$, which helps to converge the equilibrium solution (the initial state) with smaller cutoff dimensions $\chi$.
We have confirmed that the single-particle and HHG spectra shown below are hardly affected by this field through the comparison with the results for $h_z=10^{-4}v$.
The laser excitation is incorporated via the Peierls phase of the hopping parameter and $A(t)$ represents the vector potential.
Note that we set the lattice constant and the electron charge to unity. The electric field $E(t)$ is equal to $-\partial_t A(t)$.
We choose $A(t)= E_0/\Omega e^{-(t-t_0)^2/2\sigma^2} \sin[\Omega(t-t_0)]$ with $E_0$ the amplitude of the electric field.
For the following discussion, we define $\hH_{\rm kin}(t) = -v\sum_{i,\sigma} [e^{-iA(t)} \hc^\dagger_{i,\sigma} \hc_{i+1,\sigma} + h.c.]$  and $\hH_U = U\sum_i  \hn_{i,\uparrow}\hn_{i,\downarrow}$.

To discuss the HHG in the system, we introduce the current operator as 
\eqq{
\hat{j}(t) = iv\sum_\sigma \Bigl[e^{iA(t)} \hc^\dagger_{i+1,\sigma} \hc_{i,\sigma} - e^{-iA(t)}  \hc^\dagger_{i,\sigma} \hc_{i+1,\sigma}  \Bigl].
}
We evaluate the HHG spectrum as 
\eqq{
I_{\rm HHG}(\omega) = |\omega j(\omega)|^2,
}
assuming that the emitted radiation originates from the acceleration of charges by the external field.\cite{Kemper2013b}
Since the numerical simulations are restricted to a finite time range $[0,t_{\rm max}]$,
it is useful to introduce a Gaussian window $F_{\rm gauss}(t) = \exp(-\frac{(t-t_0)^2}{2\sigma^{'2}})$, which is wide enough but shorter than $t_{\rm max}$.
Then we calculate the Fourier transform of $F_{\rm gauss}(t)j(t)$ to obtain $j(\omega)$.
This allows us to suppress artificial oscillations from the sudden cut of $j(t)$ at $t=t_{\rm max}$.

The nonequilibrium dynamics of the Hubbard model is simulated with the iTEBD \cite{Vidal2007PRL,Murakami2020}
and the exact diagonalization (ED) methods.
In the implementation of iTEBD, we make use of the conservation of the number of spin-up and spin-down electrons, the 4th order Trotter-Suzuki decomposition and the 4th order commutator-free matrix exponential approximation.\cite{Alvermann2011} 
In iTEBD, we prepare the initial state with the cutoff dimension $\chi=400$, and compute the time evolution with $\chi=1000$, which is
sufficient for obtaining converged results.
In the following, we set $v$ as the unit of energy.

In this study, we discuss the single-particle spectrum $A(p,\omega) = 
-\frac{1}{\pi} {\rm Im} G^R(p,\omega)$ of the Hubbard model, where $G^R$ is the retarded electron Green's function and $p$ is the momentum.
We evaluate the single-particle Green’s function by iTEBD using a new method. Firstly, we introduce auxiliary bath sites.
We then apply a weak-enough pulse to excite an electron from the system (the Hubbard model) to the bath and measure some nonlocal correlation between the system and the auxiliary sites. One can show that this quantity corresponds to the single-particle   Green's function of the system, as explained in detail in Appendix~\ref{sec:iTEBD_Akw}.

Similar strategies to measure the single-particle   spectrum by attaching auxiliary bath sites have been proposed previously and implemented for the density matrix renormalization group (DMRG) \cite{Feiguin2019} and ED \cite{Bohrdt2018} methods.
Although these strategies essentially mimic the photoemission experiments,
our method is different from the previous ones. The previous schemes follow the evolution of  the number of particles excited to the auxiliary baths, which is connected to $A(p,\omega)$.
On the other hand, we directly measure  a nonlocal correlation function equivalent  to $G^R(p,t)$. Secondly, in order to obtain the full $A(p,\omega)$ with the previous schemes, one needs to repeat the simulations, changing the energy levels of the bath sites or the excitation frequency, while in the method used here, only a single simulation is needed. 
We also note that our approach can be easily extended to study the spectrum in a nonequilibrium setup.

\subsection{Effective models}\label{sec:model_eff}
In order to clarify the nature of HHG in the Hubbard model,
we disentangle different processes involved.
To this end, we derive effective models in the strong coupling regime $U\gg v$
using the time-dependent Schrieffer-Wolff transformation.\cite{MacDonald1988} 
This transformation is expressed as 
\eqq{\hH'(t) = e^{i\hS(t)} \hH e^{-i\hS(t)} + i(\partial_t e^{i\hS(t)}) e^{-i\hS(t)},}
 with 
 \eqq{\hS(t) = \hS^{(1)}(t) + \hS^{(2)}(t) + \hS^{(3)}(t) \cdots,} 
where $\hS^{(i)}$ is Hermitian and of the order $O( (\frac{v}{U})^i)$. 
The term $\hS^{(i)}$ is determined recursively from the terms $\hS^{(1)},\ldots,\hS^{(i-1)}$ so that  $e^{i\hS(t)} \hH e^{-i\hS(t)}$ has no terms changing the doublon and holon number up to $O(U\cdot (\frac{v}{U})^i)$. 
In the following, we denote $e^{i\hS(t)} \hH(t) e^{-i\hS(t)}$ as $\hH_{\rm Mott}(t)$ and $i(\partial_t e^{i\hS(t)}) e^{-i\hS(t)}$ as $\hH_{\rm ex}(t)$.
The excitation (doublon-holon creation) is included in $\hH_{\rm ex}(t)$, and can be expressed in the form of $-E(t)\cdot \hat{D}(t)$.
Note that $ \hat{D}(t)$ can be regarded as the dipole moment between the upper and lower Hubbard bands, and is analogous to the dipole moment between the different bands in the semiconductor models in the length gauge\cite{Vampa2014PRL,Vampa2015PRB,Luu2016,Huttner2017} and the dipole gauge.\cite{Higuchi2014,Li2020,Murakami2020} 
Keeping the terms in $\hH_{\rm Mott}(t)$ and $\hH_{\rm ex}(t)$ up to given orders in $\frac{v}{U}$,  we obtain different effective models.
In the following, we keep terms up to $O(U\cdot (\frac{v}{U})^i)$ [$O(E_0\cdot (\frac{v}{U})^i)$] for $\hH_{\rm Mott}(t)$ [$\hH_{\rm ex}(t)$], denoting the resulting operators by $\hH_{\mathrm{Mott},i}(t)$ [$\hH_{\mathrm{ex},i}(t)$]. 
The effective model consisting of $\hH_{\mathrm{Mott},N_1}(t)$ and $\hH_{\mathrm{ex},N_2}(t)$ is expressed as $\hH_{\mathrm{eff},N_1,N_2}$.

First, to determine $ \hS^{(1)}(t)$, we separate $\hH_{\rm kin}$ into four terms depending on the dynamics of the doublons and holons:
\eqq{
\hH_{\rm kin,LHB}(t) &=  -v\sum_{\langle i,j\rangle,\sigma} e^{iA(t) r_{ij}} \hh_{i,\sigma}\hh^\dagger_{j,\sigma} , \\
\hH_{\rm kin,UHB}(t) &= -v\sum_{\langle i,j\rangle,\sigma} e^{iA(t) r_{ij}}  \hd^\dagger_{i,\sigma}\hd_{j,\sigma} ,\\
\hH_{\rm kin,+}(t) &=  -v\sum_{\langle i,j\rangle,\sigma} e^{iA(t) r_{ij}} \hd^\dagger_{i,\bar{\sigma}}\hh^\dagger_{j,\sigma}  , \\
\hH_{\rm kin,-}(t) & =   -v\sum_{\langle i,j\rangle,\sigma} e^{iA(t) r_{ij}} \hh_{i,\sigma}\hd_{j,\bar{\sigma}} ,
}
where we introduced $\hh^\dagger_{i,\sigma}\equiv \bar{n}_{i,\bar{\sigma}} \hc_{i,\sigma} $ and $\hd^\dagger_{i,\sigma}\equiv n_{i,\sigma} \hc^\dagger_{i,\bar{\sigma}}$.
$\hh^\dagger_{i,\sigma}$ ($\hd^\dagger_{i,\sigma}$) creates a holon (doublon) at site $i$ from $\hc^\dagger_{i,\sigma}|{\rm vac}\rangle$.
Note that $\hh^\dagger_{i,\sigma}$ and $\hd^\dagger_{i,\sigma}$ are not normal fermionic operators. 
 $r_{ij}$ indicates the space vector from site $j$ to site $i$ and $\langle i,j\rangle$ indicates that $i$ and $j$ are nearest neighbors.
Here, $\hH_{\rm kin,-}(t)= \hH_{\rm kin,+}^{\dagger}(t)$ and $\hH_{\rm kin}(t)=\hH_{\rm kin,LHB}(t) + \hH_{\rm kin,UHB}(t) +  \hH_{\rm kin,+}(t) + \hH_{\rm kin,-}(t)$.
$\hH_{\rm kin,LHB}(t)$ changes the position of a holon, while $\hH_{\rm kin,UHB}(t)$ change the position of a doublon. 
$\hH_{\rm kin,+}(t)$ creates a doublon-holon pair at neighboring sites, while $\hH_{\rm kin,-}(t)$ annihilates a doublon-holon pair at neighboring sites.

The component of $\hH_{\rm Mott}(t)$ of order $O(U\cdot (\frac{v}{U})^1)$ can be expressed as 
\eqq{
\hH_{\rm kin,LHB} +  \hH_{\rm kin,UHB} + \hH_{\rm kin,+} + \hH_{\rm kin,-}  + i[\hS^{(1)},\hH_{U}] .
}
Thus, we require $\hH_{\rm kin,+} + \hH_{\rm kin,-}  + i[\hS^{(1)},\hH_{U}]=0$, which is satisfied by 
 \eqq{
 \hS^{(1)} = \frac{-i}{U}[\hH_{\rm kin,+} - \hH_{\rm kin,-} ].
 }
As a result, we obtain 
\eqq{
\hH_{\rm Mott,1}(t) &= \hH_{\rm kin,LHB}(t) +  \hH_{\rm kin,UHB}(t)+ \hH_U,
}
and $\hH_{\rm ex,1}(t) = -E(t)\hat{D}^{(1)}(t)$ with
\eqq{
 \hat{D}^{(1)}(t) &= \frac{v}{U} \sum_{\langle i,j\rangle,\sigma} [r_{ij} e^{iA(t)r_{ij}} \hd^\dagger_{i,\bar{\sigma}}\hh^\dagger_{j,\sigma} + h.c.] .
 }

The lowest-order effective model $\hH_{\rm eff,1,1}=\hH_{\rm Mott,1}(t) 
+ \hH_{\rm ex,1}(t)$ has a direct formal correspondence with the semiconductor models in the dipole gauge.\cite{Higuchi2014,Li2020,Murakami2020}
Namely, $\hH_{\rm kin,LHB}(t)$ corresponds to the kinetic term for the valence band electrons with the intraband acceleration, $\hH_{\rm kin,UHB}(t)$ to the analogous term for the conduction band, $\hH_U$ corresponds 
to the difference of the band energy, and $\hat{D}^{(1)}(t)$ corresponds to the interband dipole moment in the semiconductor models, see Eq.~\eqref{eq:semiconductor} in Appendix~\ref{sec:semicon}.

Next, we consider the higher-order corrections.
The higher order terms of $\hS$ are iteratively determined
as shown in Appendix~\ref{sec:Appendix_eff},
which yields the expressions for $\hH_{\mathrm{Mott},N_1}(t)$ and $\hH_{\mathrm{ex},N_2}(t)$ with $N_1>1$ and $N_2>1$.
For example, the $O(U\cdot (\frac{v}{U})^2)$ component in $\hH_{\mathrm{Mott}}(t)$, i.e. $\hH_{\mathrm{Mott},2}(t)-\hH_{\mathrm{Mott},1}(t)$, can be written as 
\eqq{
&\hH^{(2)}_{\rm kin,LHB} +  \hH^{(2)}_{\rm kin,UHB} + \hH^{(2)}_{U,\rm{shift}} \nonumber\\
       &\;\;\;+  \hH_{\rm spin,ex} +  \hH_{\rm dh,ex} +  \hH^{(2)}_{\rm dh,slide}. \label{eq:detail}
}
Here, 
$\hH^{(2)}_{\rm kin,LHB}$ ($\hH^{(2)}_{\rm kin,UHB}$) describes the correction to the holon (doublon) hopping, which includes the 2nd neighbor hopping.
$ \hH^{(2)}_{U,\rm{shift}}$ describes the shift of the local interaction $U$, $\hH_{\rm dh,ex}$ is the exchange coupling of the doublon and holon, and $\hH^{(2)}_{\rm dh,slide}$ describes the simultaneous hopping of a doublon and a holon to the neighboring sites.
The expressions for these terms are given in Appendix~\ref{sec:Appendix_eff}.
In particular, 
\eqq{
\hH_{\rm spin,ex}= J_{\rm ex} \sum_i \hat{\bf  s}_i\cdot \hat{\bf  s}_{i+1} \label{eq:H_spin}
}
describes the spin exchange, where $J_{\rm ex} = \frac{4v^2}{U}$, $\hat{\bf s}_i = \frac{1}{2}\sum_{\alpha,\beta} \hc^\dagger_{i,\alpha} \boldsymbol{\sigma}_{\alpha,\beta} \hc_{i,\beta}$,
and $\boldsymbol{\sigma}$ denotes the Pauli matrices.
It is noteworthy that the spin exchange term does not have any direct correspondence in the semiconductor systems.
Also note that $\hH_{\rm Mott,2}(t)$ becomes the well-known $t$-$J$ model 
after subtracting several terms irrelevant for a small number of holes.
In general, with increasing $N_1$, $\hH_{\mathrm{Mott},N_1}(t)$ starts to develop $N_1$-th neighbor (correlated) hoppings of doublons and holons.
As for $\hH_{\mathrm{ex},N_2}(t)$, the higher order corrections to 
$\hat{D}(t)$ include the creation of doublon-holon pairs on $N_2$-th 
nearest sites although the coefficient becomes smaller.
In the following discussion,
we find that this long-ranged dipole moment is important for the HHG process.

In the evaluation of the current in the effective model, we compute the expectation value of $\hat{j}(t)$.
Strictly speaking, under the unitary transformation, the current operators changes as  $e^{i\hS(t)}\hat{j}(t)e^{-i\hS(t)}=\hat{j}(t)  + \mathcal{O}{(\frac{v}{U})}$.
However, it turns out that the $\mathcal{O}{(\frac{v}{U})}$ correction has minor effects on the HHG spectrum in the parameter region considered here ($U=10$), where its shape, cutoff energy and  intensity are hardly affected.

 %%%%%%%%%%%%%%%%%%%%%%%%%%%%%%%%%%%%%%%%%%%%%
 \begin{figure}[t]
  \centering
    \hspace{-0.cm}
    \vspace{0.0cm}
\includegraphics[width=60mm]{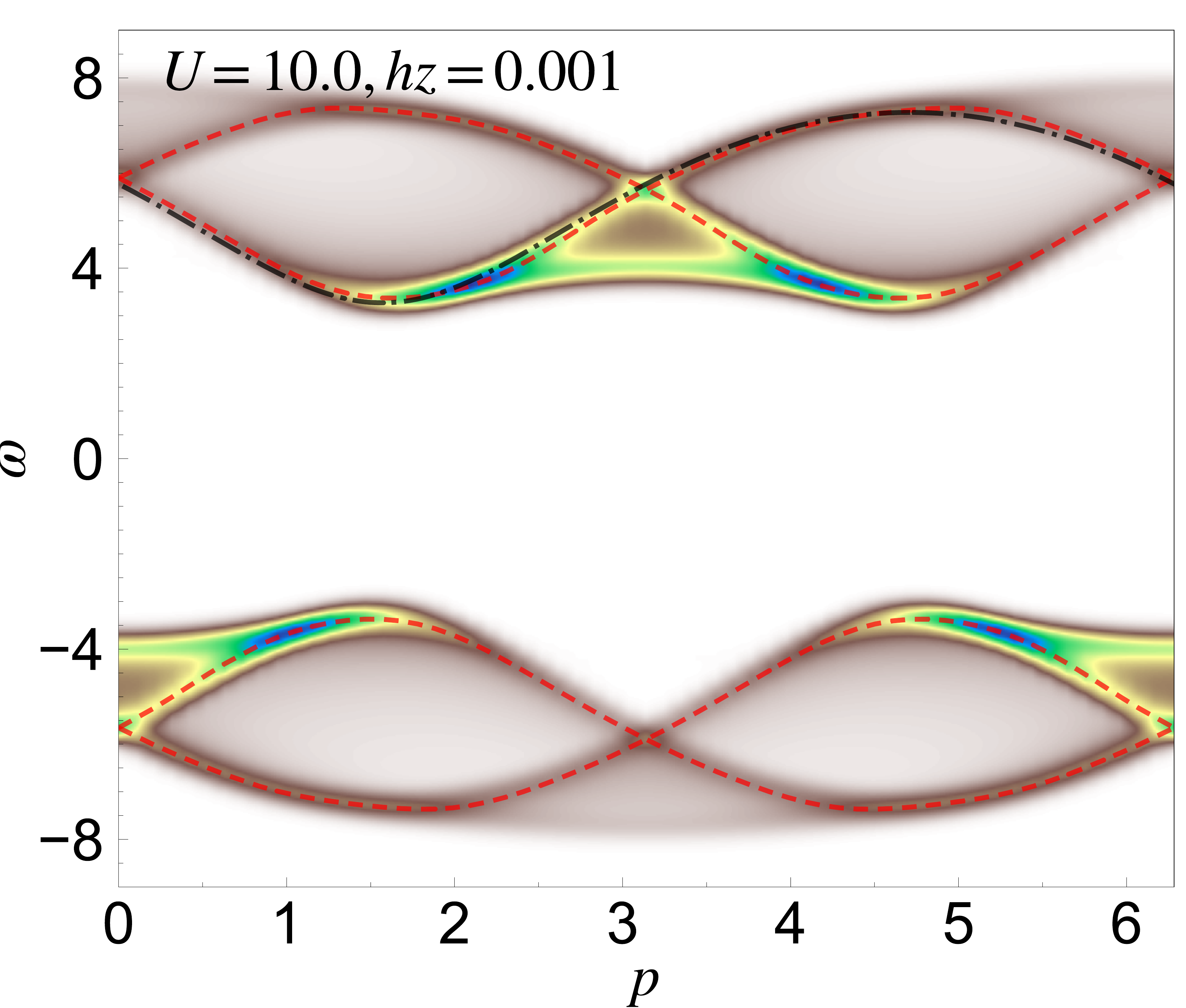} 
  \caption{Single-particle spectrum $A(p,\omega)$ in equilibrium calculated by iTEBD for $U=10$ and $h_z=0.001$. The red dashed lines indicate the peak positions of the spectrum at each $p$. The black dot-dashed line indicates $\epsilon_d(p_d)$ (Eq.~\eqref{eq:Bethe}). To obtain $A(p,\omega)$ from $G^R(p,t)$, we apply a Gaussian window $\exp(-\frac{t^2}{2\sigma^{'2}})$ with $\sigma'=5.0$ in the Fourier transformation. }
  \label{fig:fig1}
\end{figure}
%%%%%%%%%%%%%%%%%%%%%%%%%%%%%%%%%%%%%%%%%%

\section{Results}\label{sec:results}
In the following, we mainly consider systems deep in the Mott insulating phase ($U=10$).
We use pump frequencies $\Omega=0.5$, $0.75$ much 
smaller than the Mott gap to observe many high-harmonic peaks. Small pump
frequencies are usually used in the experiments in semiconductors to avoid damaging the sample material.
The center and width of the applied pulse are set to $t_0=60$ and $\sigma=15$, respectively. 
The convergence with respect to the cutoff dimension 
$\chi$ becomes generally worse for smaller $U$ due to the smaller gap.

\subsection{Doublon/holon dispersion and single-particle spectrum} \label{sec:spectra} 

Before studying HHG, we first explain the nature of the elementary excitations in the 1d Hubbard model and the single-particle spectrum. This will be helpful for the following analysis.
The 1d Hubbard model in equilibrium can be solved exactly using the Bethe ansatz.\cite{Essler2005,Oka2012PRB}
At half filling, there exist two types of elementary exictations: (i) gapped spinless excitations called holons and anitiholons (doublons)
and (ii) gapless charge neutral excitations called spinons. Physical excitations are constructed from these elementary excitations.
In particular, a holon (doublon) is parametrized by a quantity called rapidity $k$, where the corresponding momentum $p_h(k)$ ($p_d(k)$) and the energy $\epsilon_h(k)$ ($\epsilon_d(k)$)
are given by
\begin{subequations}\label{eq:Bethe}
\eqq{
p_h(k) &= p_d(k) + \pi = \frac{\pi}{2}-k-2\int_0^\infty \frac{d\omega}{\omega} \frac{\mathcal{J}_0(\omega) \sin(\omega\sin k)}{1+\exp(\frac{U\omega}{2})}, \\
\epsilon_h(k) &= \epsilon_d(k)  \nonumber \\
&= \frac{U}{2}+2\cos k+2\int_0^\infty \frac{d\omega}{\omega} \frac{\mathcal{J}_1(\omega) \cos(\omega\sin k)e^{-\frac{U\omega}{4}}}{\cosh(\frac{U\omega}{4})},
}
\end{subequations}
respectively. Here $\mathcal{J}_n$ is the Bessel function.
We shall see that this dispersion describes the kinetics of a doublon and a holon under strong fields and is thus related to HHG, see Sec.~\ref{sec:subsycle}.
 %%%%%%%%%%%%%%%%%%%%%%%%%%%%%%%%%%%%%%%%%%%%%
 \begin{figure}[t]
  \centering
    \hspace{-0.cm}
    \vspace{0.0cm}
 \includegraphics[width=65mm]{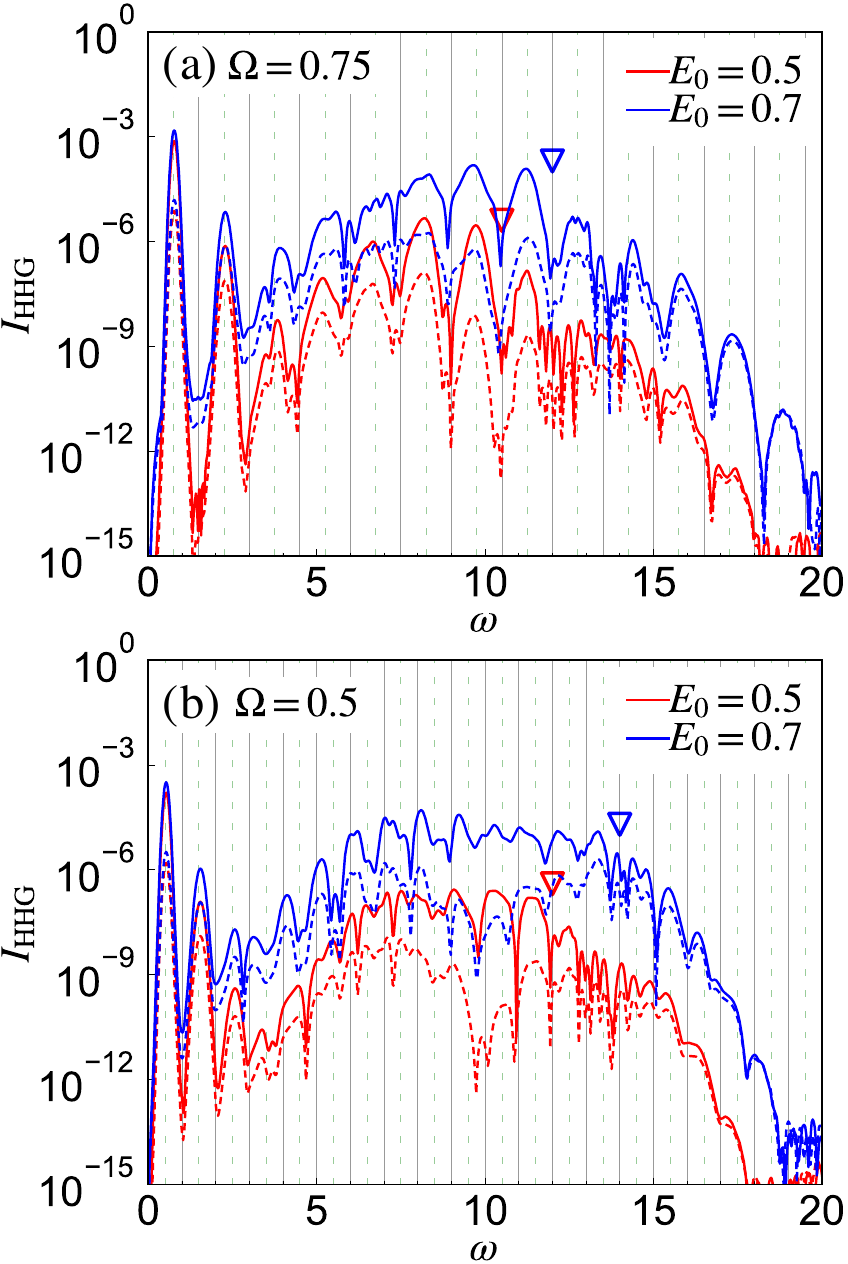} 
  \caption{
HHG spectra of the Mott insulator for (a) 
  $\Omega=0.75$ and (b) $\Omega=0.5$ at field strengths $E_{0}=0.5$ and 0.7.
  Solid lines indicate the full HHG spectra, while the dashed lines indicate the contribution from the doublon-holon hopping $I_{\rm hop}$.
    Here, the Mott gap and the maximum band-energy difference extracted from the single-particle spectrum are $6.7$ and $15.0$, respectively.
 Inverted triangles indicate the cutoff frequency determined by the criterion  [\onlinecite{Note1}].
 The solid vertical lines correspond to even harmonics of $\Omega$, while the dashed ones show odd harmonics. 
  The model parameters are $U=10$ and $h_z=0.001$.
  The pulse parameters are $t_0=60$, and $\sigma=15$, while $\sigma'=20$ is used for the Fourier transformation of the current.}
  \label{fig:fig2}
\end{figure}
%%%%%%%%%%%%%%%%%%%%%%%%%%%%%%%%%%%%%%%%%%
 %%%%%%%%%%%%%%%%%%%%%%%%%%%%%%%%%%%%%%%%%%%%%
 \begin{figure}[t]
  \centering
    \hspace{-0.cm}
    \vspace{-0.0cm}
 \includegraphics[width=65mm]{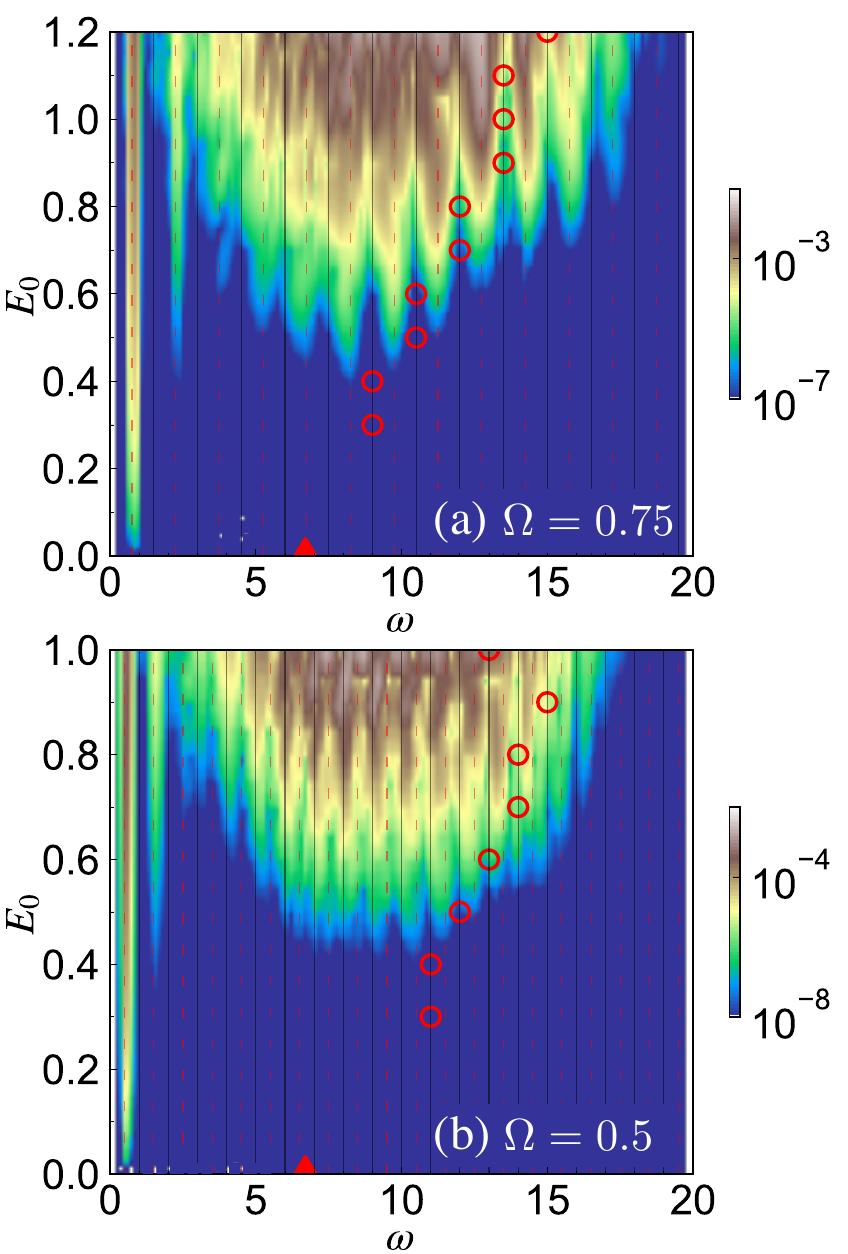} 
  \caption{HHG spectra in the plane of frequency $\omega$ and field strength $E_0$ for (a) $\Omega=0.75$ and (b) $\Omega=0.5$.  The red circles indicate the cutoff frequencies and the filled red triangle shows the Mott gap. 
  The solid vertical lines correspond to even harmonics of $\Omega$, while the dashed ones show odd harmonics.   The model parameters are $U=10$ and $h_z=0.001$.
  The pulse parameters are $t_0=60$, and $\sigma=15$, while $\sigma'=20$ is used for the Fourier transformation of the current.
}
  \label{fig:fig2_2}
\end{figure}
%%%%%%%%%%%%%%%%%%%%%%%%%%%%%%%%%%%%%%%%%%

We show how these many-body elementary excitations are reflected in the single-particle spectrum $A(p,\omega)$, see Fig.~\ref{fig:fig1}.
The single-particle excitation consists of holons, spinons, and doublons and as a result it exhibits multiple bands \cite{Jeckelmann2007} with a direct gap near $p=\pi/2$. 
In particular, one can see that the peaks in the spectrum (red dashed lines) match well with the doublon energy obtained from the Bethe ansatz (Eq.~\eqref{eq:Bethe}), see the black line.
Note that this match is nontrivial since the single-particle excitation consists of combinations of holons, spinons, and doublons\cite{Bohrdt2018} and the weight for each state is not a priori clear.
The structure of the Hubbard bands is qualitatively different from that obtained by the Hubbard-I approximation\cite{Hubbard_1} or the dynamical mean-field theory,\cite{Murakami2018PRL} which gives dispersions of 
the upper and lower Hubbard bands that are parallel. 
The Mott gap $\Delta_{\mathrm{Mott}}$ estimated from the single-particle spectrum is about $6.7$.

\subsection{HHG spectra}\label{sec:hhg}

Here, we study the HHG in the 1d Mott insulator.
In Figs.~\ref{fig:fig2}(a) and \ref{fig:fig2}(b), we show the HHG 
spectra for $\Omega=0.75$ and $\Omega=0.5$, respectively, with 
$E_{0}=0.5$ and 0.7.
For $\Omega=0.75$, we see clear intensity peaks at the odd harmonics, as is expected in the time-periodic steady states of a system with inversion symmetry.\cite{Cohen2019NatCom}
The HHG intensity decreases monotonically for energies below the Mott gap, while above it, there exists a HHG plateau. 
The cutoff of the plateau, which we extract as described in [\onlinecite{Note1}], increases with increasing field intensity.
For $\Omega=0.5$, we see peaks in $I_{\mathrm{HHG}}(\omega)$, but they are not necessarily located at the odd harmonics, in particular for $\omega\simeq\Delta_{\rm Mott}$.
Interestingly, one can observe peaks at even harmonics, which is unexpected in the present system with inversion symmetry.
We believe that this is because the pulse is not long enough, and some potentially relevant dephasing channels (such as electron-phonon couplings) are missing in the present model, so that the system does not reach a time-periodic steady state. 
Indeed, a standard semiconductor system with a similar band gap and without dephasing also shows a HHG spectrum with some peaks deviating from the odd harmonics for the same pump pulse (see Appendix~\ref{sec:semicon}).
On the other hand, with a longer pulse and a phenomenological dephasing, we obtain clearer peaks at the odd harmonics.
Figures~\ref{fig:fig2_2} (a) and \ref{fig:fig2_2}(b) show the color plot of 
the HHG intensity $I_{\rm HHG}(\omega)$ in the plane of the frequency 
($\omega$) and the field strength ($E_0$), and the circles represent the 
cutoff frequencies of the plateau.\footnote{The criterion for the 
determination of cutoff frequencies is as follows. We pick up the 
maximum of $I_{\rm HHG}(\omega)$ in the interval of $\omega\in[2n\Omega, 
(2n+2)\Omega]$ for each $n$, which we represent $I_{n}$. Then, if $I_{n+1}/I_{n}<\alpha=0.35$, we set the cutoff frequency to $(2n+2)\Omega$. This criterion is consistent with Ref.~\onlinecite{Murakami2018PRL}}
In contrast to the case of atomic gases,~\cite{Corkum1993PRL,Lewenstein1994}
the cutoff frequencies do not scale as $E_0^2$. 
Rather, our results indicate a linear scaling $\omega_{\rm cut} \simeq 
\Delta_{\rm Mott} + \alpha(\Omega) E_0$ as in the case of 
semiconductors.\cite{Vampa2015PRB} This behavior is consistent 
with the previous DMFT study of the higher dimensional Hubbard model.\cite{Murakami2018PRL}
$\alpha(\Omega)$ decreases with increasing excitation frequency $\Omega$, and it roughly scales as $1/\Omega$.

Now, we discuss the origin of HHG and analyze the underlying processes in detail. We first identify the different contributions to the HHG spectrum. The current in semiconductors can be classified into an interband current (particle-hole annihilation/creation) and an intraband current. 
In the Hubbard model, we analogously consider the current originating 
from the annihilation/creation of a doublon and holon $j_{\rm ac}$ and 
the current associated with hopping of doublons or holons $j_{\rm hop}$, which conserves the number of these carriers.
The operators corresponding to these currents can be expressed as 
\eqq{
\hat{j}_{\rm ac}(t) &= iv\sum_{\langle i j \rangle \sigma}r_{i,j}e^{iA(t)r_{ij}}  \Bigl[ \hd^\dagger_{i,\sigma} \hh^\dagger_{j,\bar{\sigma}}  + \hh_{i,\sigma} \hd_{j,\bar{\sigma}}   \Bigl],\nonumber\\
\hat{j}_{\rm hop}(t) &= iv\sum_{\langle i j \rangle \sigma}r_{i,j}e^{iA(t)r_{ij}}  \Bigl[ \hh_{i,\sigma} \hh^\dagger_{j,\sigma}  + \hd^\dagger_{i,\sigma} \hd_{j,\sigma}   \Bigl].
}
We denote the contribution to HHG from these currents by $I_{\rm ac}(\omega)=|\omega j_{\rm ac}(\omega)|^2$ and $I_{\rm hop}(\omega)=|\omega j_{\rm hop}(\omega)|^2$.
The evaluation of these contributions shows that $I_{\rm ac}(\omega)$ 
dominates over $I_{\rm hop}(\omega)$, in particular around the cutoff 
frequencies [Figs.~\ref{fig:fig2}(a) and \ref{fig:fig2}(b)].
Therefore the annihilation of doublon-holon pairs is the dominant source of HHG in large-gap Mott insulators, which is consistent with the conclusion from a previous DMFT analysis.\cite{Murakami2018PRL}
Thus, one can expect that he frequency range of the plateau approximately corresponds to the possible energies of the doublon-holon pairs (in the presence of the external field) just before the recombination.

\subsection{Subcycle analysis}\label{sec:subsycle}

\begin{figure}
  \centering
    \hspace{-0.cm}
    \vspace{0.0cm}
  \includegraphics[width=87mm]{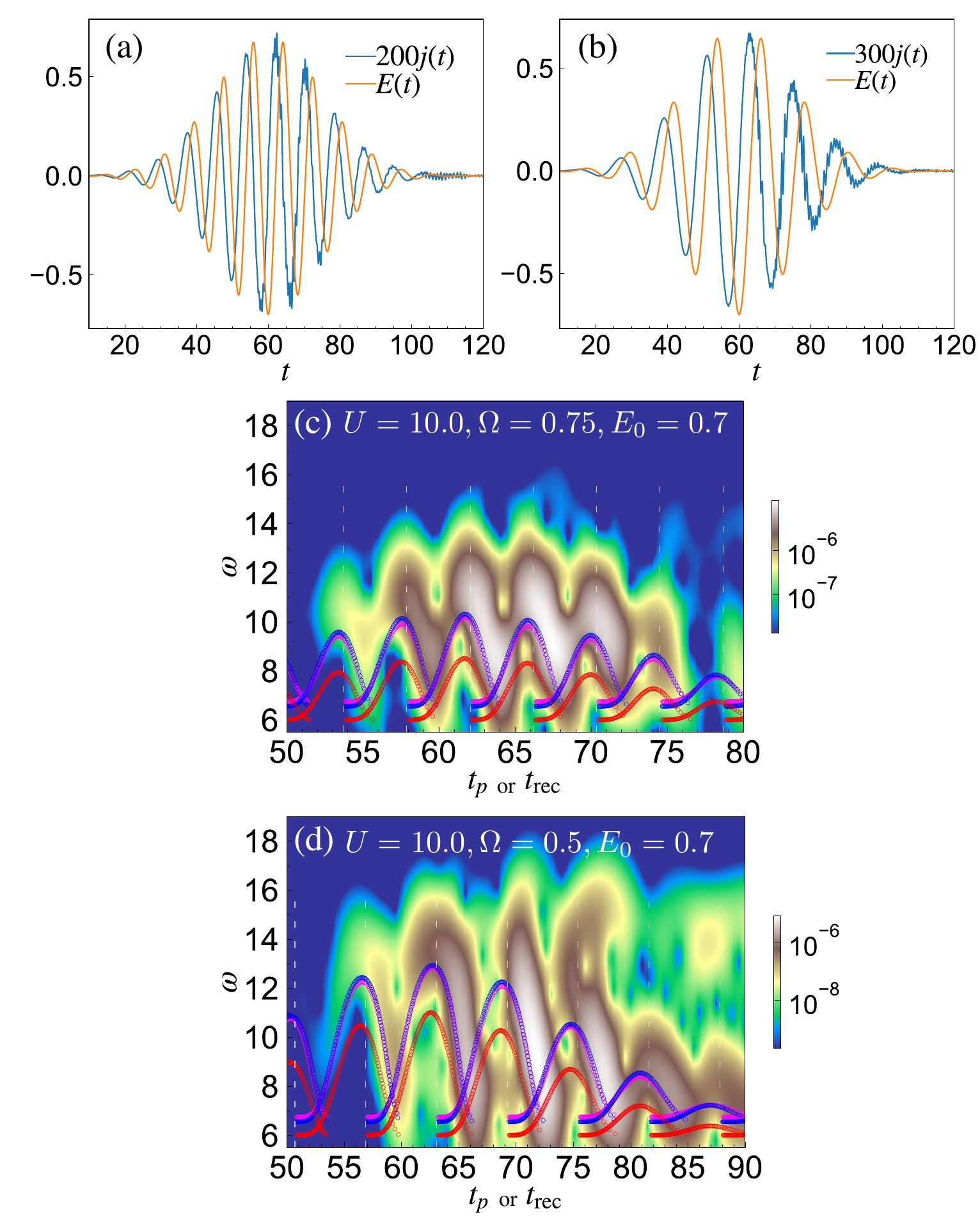} 
  \caption{(a,b) Applied electric field and induced current for (a) $\Omega=0.75$ and (b) $\Omega=0.5$. (c,d)  Subcycle analysis of the HHG, $I_{\rm HHG}(\omega,t_p)$, for (c) $\Omega=0.75$ and (d) $\Omega=0.5$.
The blue markers show the results of the semiclassical analysis ($\omega_{\rm emit}(t_{\rm rec})$) with $\epsilon_g(p)$ obtained from the Bethe ansatz results Eq.~\eqref{eq:Bethe},
while the red markers correspond to those for the unrenormalized dispersion, $U-4v \sin(p)$.
 The pink markers are the results obtained using the dispersion extracted from the single-particle spectrum, i.e. Fig.~\ref{fig:fig1}, and they almost overlap with the blue markers. 
The vertical dashed lines indicate the times when $E(t)=0$. The parameters of the system and the pump field are the same as in Fig.~\ref{fig:fig2}. 
 }
  \label{fig:fig3}
\end{figure}

Next, we address  the dynamics of the doublons and holons during the pulse. To gain insights into this, we perform a subcycle analysis of the induced current $j(t)$ (see  Figs.~\ref{fig:fig3}(a) and \ref{fig:fig3}(b) ) by applying a short window function, $F_{\rm window}$, around $t_p$.
Specifically, we consider a windowed Fourier transform $j(\omega,t_p) = \int dt e^{i\omega t}  F_{\rm window}(t-t_p)j(t)$ and evaluate $I_{\rm HHG}(\omega,t_p)\equiv|\omega j(\omega,t_p)|^2$.
The latter function provides the time-resolved spectral features of the emitted light around $t_p$. In the following, we choose $F_{\rm window}(t)=\exp(-\frac{t^{2}}{2{\sigma'}^{2}})$ and $\sigma'=0.8$.
In Figs.~\ref{fig:fig3}(c) and \ref{fig:fig3}(d), we plot $I_{\rm HHG}(\omega,t_p)$ for the Mott insulator. 
As in the case of semiconductors, the high frequency light is emitted around $E(t_p)=0$, i.e., where $|A(t)|$ is maximum (see Appendix~\ref{sec:semicon} for semiconductor data).

 %%%%%%%%%%%%%%%%%%%%%%%%%%%%%%%%%%%%%%%%%%%%%
%%%%%%%%%%%%%%%%%%%%%%%%%%%%%%%%%%%%%%%%%%

In semiconductors, the features of $I_{\rm HHG}(\omega,t_p)$ can be explained by the three-step model formulated with the semiclassical theory for electrons and holes.\cite{Vampa2014PRL,Vampa2015PRB}
In this picture, (i) both of an electron and a hole are simultaneously created at the same position 
via tunneling; (ii) they move around; (iii) they return back to the original position and recombine by emitting the light.
The dynamics can be described with the equation of motion
\begin{subequations}\label{eq:semi_classic}
\eqq{
\frac{d x_{\rm rel}(t)}{dt} &= \partial_p \epsilon_{g}(p)\bigl |_{p=p(t)}, \\
p(t) &=p_0 -A(t)+A(t_0).
}
\end{subequations}
Here $x_{\rm rel}(t)$ indicates the relative distance between the electron and the hole, and $\epsilon_{g}(p)=\epsilon_{c}(p)-\epsilon_{v}(p)$ with $\epsilon_{c}(p)$ ($\epsilon_{v}(p)$) denoting the energy of the conduction (valence) band electron with momentum $p$. The velocity of an electron (hole) is given by the derivate $\partial_p\epsilon_{c}(p)$ ($\partial_p\epsilon_{v}(p)$), and the effect of the field is taken into account via the shift of the momentum.
Here, it is assumed that initially at $t_0$ the electron-hole pair is created with $p_0$ and $|x_{\rm rel} (t_0 )$| = 0, where $\epsilon_g(p)$ becomes minimum at $p_0$.
The light emission from the recombination occurs at $t_{\rm rec}$ ($>t_0$) when $|x_{\rm rel}(t_{\rm rec})|=0$.
The frequency of the emitted light is assumed to be equal to the energy of the electron-hole pair: $\omega_{\rm emit}(t_{\rm rec})=\epsilon_{g}(p(t_{\rm rec}))$.
Note that the photo-exicitation only produces electron-hole pairs with zero total momentum. $\epsilon_{g}(p)$ describes the energy of such an electron-hole pair as a function of half the relative momentum, $p$. Equation~\eqref{eq:semi_classic} implies that the relative momentum is changed by the vector potential, and the dispersion relation $\epsilon_{g}(p)$ dictates the relative motion of the electron and the hole.

 %%%%%%%%%%%%%%%%%%%%%%%%%%%%%%%%%%%%%%%%%%%%%
 \begin{figure}[t]
  \centering
    \hspace{-0.cm}
    \vspace{0.0cm}
   \includegraphics[width=65mm]{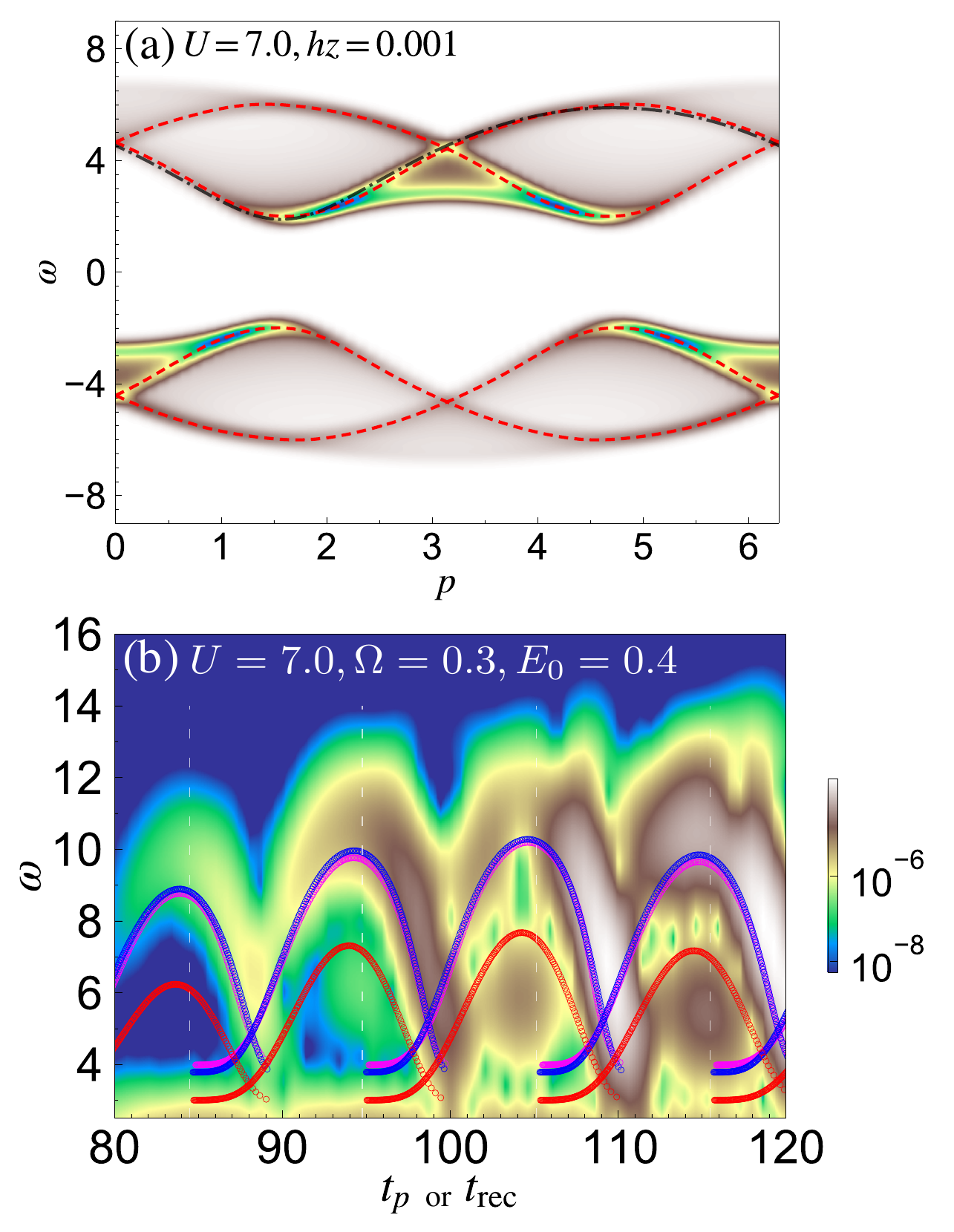} 
  \caption{(a) Single-particle spectrum $A(p,\omega)$ in equilibrium calculated by iTEBD for $U=7$ and $h_z=0.001$. The red dashed lines indicate the peak positions of the spectrum at each $p$. The black dot-dashed line indicates $\epsilon_d(p_d)$ (Eq.~\eqref{eq:Bethe}). To obtain $A(p,\omega)$ from $G^R(p,t)$, we apply a Gaussian window $\exp(-\frac{t^2}{2\sigma^{'2}})$ with $\sigma'=5.0$ for the Fourier transformation. (b) Subcycle analysis of the HHG, $I_{\rm HHG}(\omega,t_p)$ for $U=7$ and $h_z=0.001$. The parameters of the pump are $\Omega=0.3,E_0=0.4,t_0=100$ and $\sigma=30$. The blue markers show the results of the semiclassical analysis ($\omega_{\rm emit}(t_{\rm rec})$) with $\epsilon_g(p)$ obtained from the Bethe ansatz results Eq.~\eqref{eq:Bethe},
while the red markers correspond to those for the unrenormalized dispersion, $U-4v \sin(p)$.
 The pink markers are the results obtained using the dispersion extracted from the single-particle spectrum in panel (a), and they almost overlap with the blue markers.
  The vertical dashed lines indicate the times when $E(t)=0$.
 }
  \label{fig:fig3_2}
\end{figure}
%%%%%%%%%%%%%%%%%%%%%%%%%%%%%%%%%%%%%%%%%%

In the 1d Mott insulator, the relevant excitation process is the creation of doublon-holon pairs with zero total momentum $p_{\rm d}+p_{\rm h}=0$, see Refs.~\onlinecite{Oka2012PRB,Zala2012PRL}.
Such a doublon-holon pair is parametrized by half of the relative momentum between the two charge carriers $p_{\rm rel}=(p_{\rm d}-p_{\rm h})/2$, and, under an adiabatic change of the vector potential potential $A$, the relative momentum is changed to $p_{\rm rel}-A$.\cite{Oka2012PRB,Zala2012PRL}
This is the same situation as for the electron-hole pair in the semiconductor. Hence, we expect that the three-step model can be extended to the Mott insulator by considering the dispersion of a doublon-holon pair of zero total momentum with respect to the relative momentum.
In other words, the analogy suggests that the kinetics of the relative position between the doublon and the holon under strong fields is determined by this doublon-holon pair dispersion.

We test this idea using the dispersion relation of the doublon-holon pair, $\epsilon_g(p_{\rm rel}) \equiv \epsilon_d(p_{\rm rel}) +  \epsilon_h(-p_{\rm rel})$, obtained from the Bethe ansatz solution Eq.~\eqref{eq:Bethe} and substituting it into Eq.~\eqref{eq:semi_classic}. The process of computing $t_{\rm rec}$ for given $t_0$ is the same as in the semiconductor case.\footnote{To be more precise, the initial condition is assumed to be $|x_{\rm rel}(t_0)|=1$, i.e., the pair is created at neighboring sites. The light emission from the recombination occurs at $t_{\rm rec}$ ($>t_0$) when $|x_{\rm rel}(t_{\rm rec}))|=1$ is realized.}
The resulting $\omega_{\rm emit}(t_{\rm rec})$ is shown by blue markers in Figs.~\ref{fig:fig3}(c) and \ref{fig:fig3}(d).
\footnote{The markers consist of separated parts since for some $t_0$ a doublon and a holon never recombine. Also for some $t_{\rm rec}$, two values of $\omega_{\rm emit}(t_{\rm rec})$ exist since two different pairs created at different times recombine at the same time $t_{\rm rec}$ }
One can see that the semiclassical analysis explains the fact that the 
high-frequency light is emitted when $A(t)\simeq 0$, and the blue 
markers approximately overlap with the strong intensity region in $I_{\rm HHG}(\omega,t_p)$.
In general, the semiclassical analysis tends to underestimate the frequencies where the maxima of $I_{\rm HHG}(\omega,t_p)$ are located.
The match between $I_{\rm HHG}(\omega,t_p)$ and the semiclassical result
is worse for higher frequency excitations as seen from the comparison between $\Omega=0.75$ and $\Omega=0.5$.
Such disagreement is expected because the tunneling picture becomes worse for high frequency excitations.
In fact, the agreement between $I_{\rm HHG}(\omega,t_p)$ and the semiclassical result becomes better for smaller $U$ and $\Omega$ (Fig.~\ref{fig:fig3_2}).
We note that the agreement with the semiclassical analysis is as good as that between the standard semiconductor system and the corresponding semiclassical analysis, see Fig.~\ref{fig:fig_A1} in Appendix~\ref{sec:semicon}.
Hence, we conclude that the present semiclassical analysis for the 1d Mott insulator captures important aspects of the doublon-holon dynamics and their recombination.

 %%%%%%%%%%%%%%%%%%%%%%%%%%%%%%%%%%%%%%%%%%%%%
 \begin{figure}
  \centering
    \hspace{-0.cm}
    \vspace{0.0cm}
  \includegraphics[width=68mm]{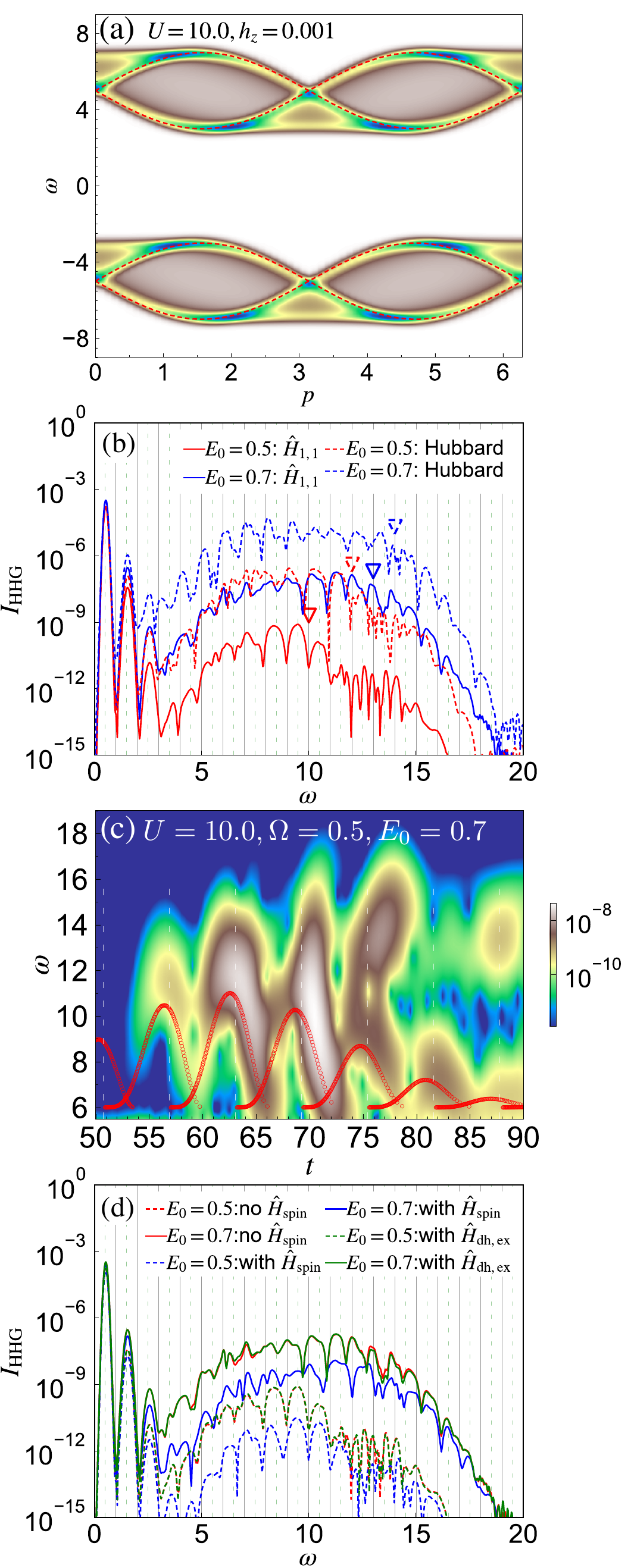} 
  \caption{(a) Single-particle  spectrum $A(p,\omega)$ of the effective 
  model $\hH_{\rm Mott,1}$ in equilibrium evaluated with iTEBD for $U=10$. 
  Here, the spin configuration of the equilibrium state is determined by $\hH_{\rm spin} = \hH_{\rm spin,ex}+h_z \sum_i (-)^i \hs_{z,i} $ with $h_z=0.001$. The red lines indicate $\pm \frac{U}{2} \pm 2v \sin(p)$.  (b) Comparison of the HHG spectra of the Hubbard model and the effective model $\hH_{\rm eff,1,1}$ for $\Omega=0.5$. (c) 
   Subcycle analysis of the HHG, $I_{\rm HHG}(\omega,t_p)$, for $\hH_{\rm eff,1,1}$. Here, the excitation parameters are $\Omega=0.5$ and $E_0=0.7$.
 The red markers indicate the result of the semiclassical theory ($\omega_{\rm emit}(t_{\rm rec})$) with the unrenormalized dispersion, $U-4v \sin(p)$.
 (d) Comparison between the HHG spectra from $\hH_{\rm eff,1,1}$, $\hH_{\rm eff,1,1}+\hH_{\rm spin}$ and $\hH_{\rm eff,1,1}+\hH_{\rm dh,ex}$ for $\Omega=0.5$. 
 In(b)-(d), we use $U = 10$, $h_z = 0.001$, $t_0 =60$ and $\sigma=15$. Here, the Mott gap and the maximum band-energy difference extracted from the single-particle spectrum are 6.0 and 14.0, respectively. }
  \label{fig:fig4}
\end{figure}
%%%%%%%%%%%%%%%%%%%%%%%%%%%%%%%%%%%%%%%%%%

We also show the results of the semiclassical analysis using the dispersion relation of the single-particle spectrum (the red dashed lines in Figs.~\ref{fig:fig1} and \ref{fig:fig3_2}(a)),
see the pink markers in Figs.~\ref{fig:fig3}(c),~\ref{fig:fig3}(d) and ~\ref{fig:fig3_2}(b). 
More specifically, we extract the ``$\epsilon_{c}(p)$" and ``$\epsilon_{v}(p)$" in the three step model from the dispersion of the upper and lower Hubbard bands in  the single-particle spectrum.
Since they agree well with $\epsilon_{d}(p)$ and $-\epsilon_{h}(p)$, respectively, the results of this semiclassical analysis also agree well with the above results (the blue markers).
This result demonstrates that, in the present case of the Hubbard model, the dispersion obtained from the single-particle spectrum provides the relevant information on the kinetics of the doublons and holons, as in the semiconductor.
However, we emphasize that it is not necessarily the case for general SCESs.  We will come back to this point in Sec.~\ref{sec:discuss}.

In addition, we show the results of the semiclassical analysis based on  
the unrenormalized dispersion $\epsilon_{g}(p)=U - 4v \sin(p)$  in $\hH_{\rm Mott,1}$ by the red 
markers in Figs.~\ref{fig:fig3}(c), ~\ref{fig:fig3}(d) and ~\ref{fig:fig3_2}(b).
By comparing the red and blue makers, one realizes that the renormalization of the Mott gap as well as the dispersion are important for a reasonable agreement between the semiclassical results and $I_{\rm HHG}(\omega,t_p)$.

Finally, we note some major differences from the semiconductor results: i) the high-frequency signals remain even after the pulse, and ii) the recombination is enhanced during the period when $|E(t)|$ increases.
Interestingly, the latter observation indicates that the recombination 
happens more likely for doublon-holon pairs which move around for more than half a period ($\pi/\Omega$) after their creation by tunneling.

\subsection{Analysis of the effective models} \label{sec:hhg_eff}
In this section, we study to what extent the effective models explain the HHG in the 1d Mott insulator and clarify the role of different processes.
In Fig.~\ref{fig:fig4}(a), we show the single-particle spectrum obtained by the iTEBD for the lowest order model $\hH_{\rm eff,1,1}(t)$.
The equilibrium spin configuration is determined from the ground state of $\hH_{\rm spin} \equiv \hH_{\rm spin,ex}+h_z \sum_i (-)^i \hs_{z,i} $.
In this model, the dispersion of the upper and lower Hubbard bands matches well with $\pm\frac{U}{2}\pm 2v\sin(p)$, see Fig.~\ref{fig:fig4}(a).
Note that this dispersion ($\pm 2v\sin(p)$) is exactly that of the doublon (antiholon) and holon obtained from the Bethe ansatz for $U\rightarrow\infty$\cite{Essler2005} or from the direct construction of the wave functions.\cite{Koch1997}
In Fig.~\ref{fig:fig4}(b), we show the corresponding HHG spectrum.
The HHG intensity and the cutoff are underestimated compared to the full simulation (dashed lines), while the response at low frequencies around $\omega=\Omega$ is already well described.
The subcycle analysis for the effective model shows that the transient signal is consistent with the semiclassical analysis with the dispersion $\epsilon_g(p)=U-4v \sin(p)$, see Fig.~\ref{fig:fig4}(c).
This observation again underpins that the dispersion of the doublon-holon pair with respect to their relative momentum is closely related to the HHG in the 1d Mott insulator.
Furthermore, the effective model already captures the characteristic subcycle 
feature of the Mott HHG, points i) and ii) mentioned in the last part of 
the previous subsection.

Now, we study the effects of the higher order terms on the HHG.
First, we discuss the effect of the Heisenberg term $\hH_{\rm spin}$ on the HHG signal.
In Fig.~\ref{fig:fig4}(d), we compare the results for  $\hH_{\rm eff,1,1}(t) $ and for $\hH_{\rm eff,1,1}(t) + \hH_{\rm spin}$.
One can see that the HHG intensity is substantially suppressed by $\hH_{\rm spin}$.
One may think that this is an unexpected effect, since $\hH_{\rm spin}$ applies only to the singly-occupied sites and it looks unrelated to the dynamics of the doublons or holons.
Moreover, unlike in higher-dimensional systems, the doublon and holon dynamics does not disturb the spin background directly (spin-charge separation). 
We think that the reduction originates from a reduced recombination probability for the doublon and holon. While the doublon and holon move around, the spin background can change through the action of $\hH_{\rm spin}$. Thus, the spin background can be substantially different when the doublon and holon return back to neighboring positions for possible recombination. 
If we denote such an excited state by $|\Psi_{\rm ex}\rangle$ and the ground state by $|\Psi_0\rangle$,  the matrix element $\langle \Psi_{\rm ex}|\hat{j}|\Psi_0\rangle $ should be decreased compared to the case without $\hH_{\rm spin}$ due to the mismatch of the spin background.
This result shows that the HHG intensity in Mott systems is sensitive to spin dynamics, which raises interesting questions about the effects of different types of spin couplings and external magnetic fields on the HHG in Mott insulators.
In a similar way, one can study the effects of the exchange term for the doublon and holon ($\hH_{\rm dh,ex}$), which turn out to be minor, see Fig.~\ref{fig:fig4}(d).

 %%%%%%%%%%%%%%%%%%%%%%%%%%%%%%%%%%%%%%%%%%%%%
 \begin{figure}
  \centering
    \hspace{-0.cm}
    \vspace{0.0cm}
  \includegraphics[width=58mm]{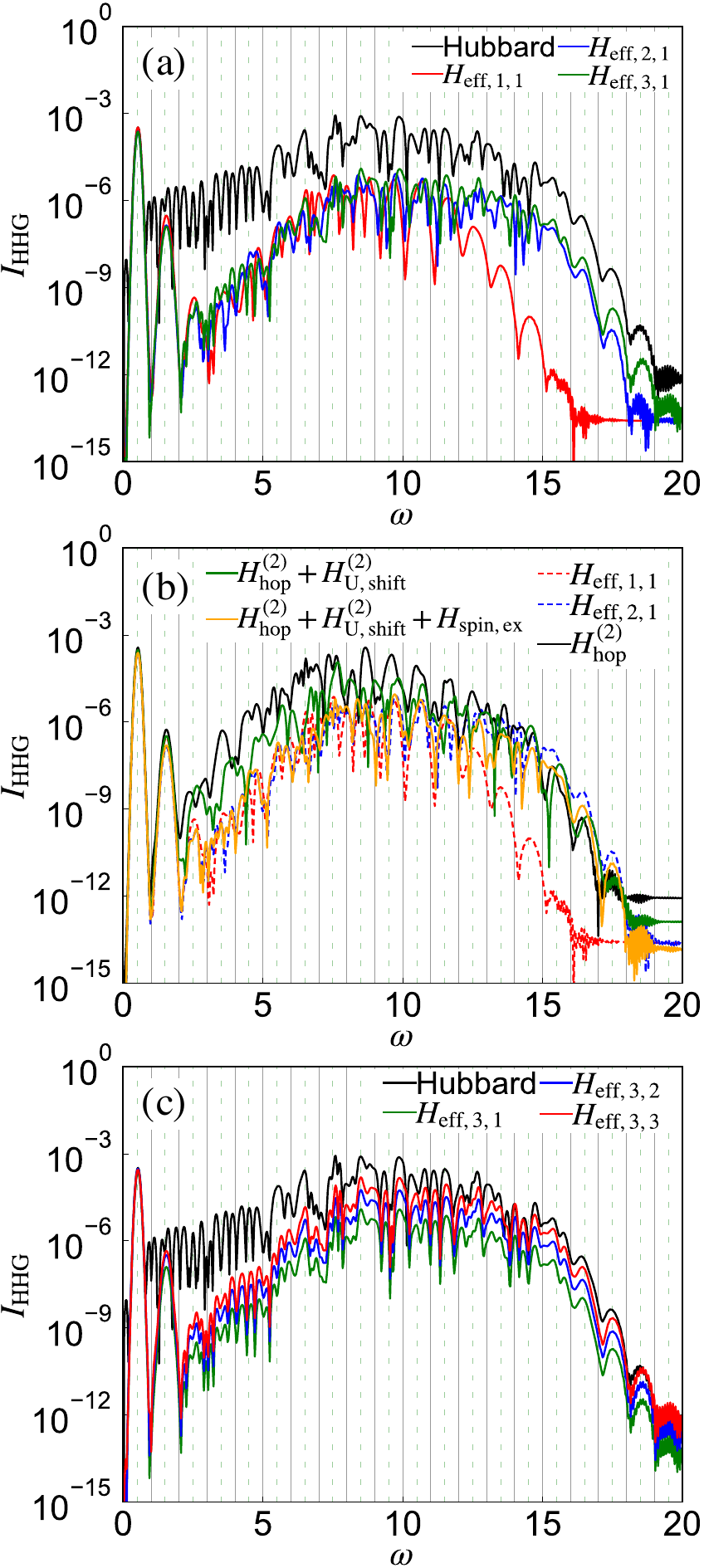} 
  \caption{HHG spectra for the effective models evaluated with ED for $U=10,h_z=0$ and $L=10$. The corresponding Hamiltonians are indicated in the labels. In particular, the labels for the solid lines in panel (b) indicate the correction added to $H_{\rm eff,1,1}$. For example, ``$H_{\rm hop}^{(2)}$" means that $\hH_{\rm eff,1,1}+\hH_{\rm hop}^{(2)}$  is used for the simulation. The parameters for the pump pulse are $\Omega=0.5$, $E_0=0.7$, $t_0=60$ and $\sigma=15$. }
  \label{fig:fig5}
\end{figure}
%%%%%%%%%%%%%%%%%%%%%%%%%%%%%%%%%%%%%%%%%%

The role of the other higher order corrections is studied with ED, since the terms involving sites beyond the nearest neighbor are difficult to treat  with iTEBD.
First, we discuss the effects of corrections in $\hH_{\rm Mott}(t)$. 
In Fig.~\ref{fig:fig5}(a), we show the result of $\hH(t)$ (Eq.~\eqref{eq:Hubbard}), $\hH_{\rm eff,1,1}(t)$, $\hH_{\rm eff,2,1}(t)$, and $\hH_{\rm eff,3,1}(t)$ for $L=10$.
The cutoff is substantially increased from $\hH_{\rm eff,1,1}(t)$ to $ \hH_{\rm eff,2,1}(t)$, and already reproduces well the cutoff in the ED results for the original Hubbard model.
\rm{The detailed information on the effects of each term in $\hH_{\rm eff,2,1}(t)$ (Eq.~\eqref{eq:detail}) is shown in Fig.~\ref{fig:fig5}(b). As is expected, the correction to the hopping of the doublons and holons, i.e. $\hH^{(2)}_{\rm hop}\equiv \hH^{(2)}_{\rm kin,LHB} +  \hH^{(2)}_{\rm kin,UHB}$, 
increases the intensity both at high frequencies and around the gap edge as well as  the cutoff frequency compared to $\hH_{\rm eff,1,1}(t)$. 
  Further inclusion of $\hH^{(2)}_{U,{\rm shift}}$ and $\hH_{\rm spin,ex}$ reduces the HHG intensity in general.  The former is expected since $\hH^{(2)}_{U,{\rm shift}}$ increases the Mott gap, while the latter effect has already been pointed out above. }
$\hH_{\rm eff,3,1}(t)$ well reproduces the shape of the HHG spectrum of the full Hubbard model.
However, this model substantially underestimates the HHG intensity.

In Fig.~\ref{fig:fig5}(c), we show the HHG spectra for $\hH_{\rm eff,3,1}(t),  \hH_{\rm eff,3,2}(t)$, and $\hH_{\rm eff,3,3}(t)$ to illustrate the effects of the higher order dipole-like terms,\;$\hH_{\rm ex}(t)$.
Interestingly, the shape of the HHG above the band gap is hardly changed but its intensity substantially increases, while the response around $\omega=\Omega$ is only slightly affected.
This shows that although the higher order corrections to the dipole term are not relevant for the response around $\omega=\Omega$, they are necessary to quantitatively
reproduce the HHG signal above the gap.
The coefficients of the higher order corrections in $\hH_{\rm ex}$ are small, but they include long-range terms, which help the creation of doublon-holon pairs via tunneling.

\subsection{Discussion: Kinetics of doublons and holons}\label{sec:discuss}
\begin{figure}[t]
  \centering
    \hspace{-0.cm}
    \vspace{0.0cm}
   \includegraphics[width=58mm]{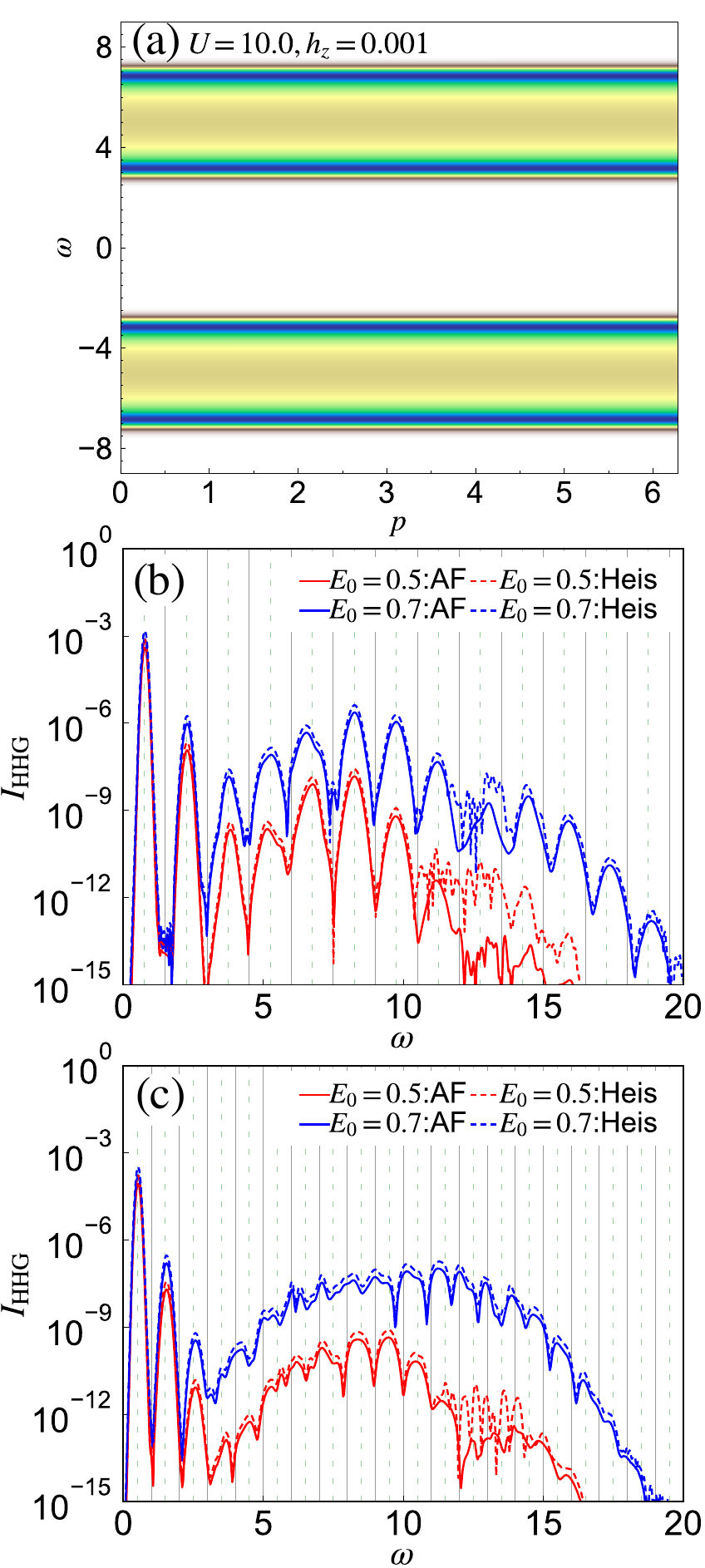} 
  \caption{(a) Single-particle  spectrum $A(p,\omega)$ of the effective model $\hH_{\rm Mott,1}$ in equilibrium evaluated with iTEBD for $U=10$. Here, the spin configuration of the equilibrium state is antiferromagnetic. (b)(c) Comparison of the HHG spectra of the effective model, $\hH_{\rm eff,1,1}$ with $U=10$, for the ground state of $\hH_{\rm spin}+h_z \sum_i (-)^i \hs_{z,i} $ and the antiferromagnetic state.
   Here, (b) is for  $\Omega=0.75$ and (c) is for $\Omega=0.5$. The parameters of the pump are $t_0=60$ and $\sigma=15$.}
  \label{fig:fig6}
\end{figure}

Here, we would like to discuss the kinetics of the doublons and holons, and its relation with the single-particle   spectrum.
First, although we showed that the band dispersion obtained from the single-particle  spectrum of the 1d Mott insulator (Fig.~\ref{fig:fig1}) can be used in a phenomenological three step model for the doublon-holon recombination, we note that the full information on the doublon and holon dynamics may not be obtainable from the single-particle spectrum in general. 
To exemplify this, we consider $\hH_{\rm eff,1,1}$ and choose the antiferromagnetic state ($|\!\uparrow,\downarrow,\uparrow,\downarrow,\cdots\rangle$) as the initial equilibrium state.
Remember that $\hH_{\rm Mott,1}$ shows degenerated ground states, i.~e., the states with no doublons and holons  become the ground states regardless of the spin configuration.
In this case, the single-particle   spectrum is independent of momentum and there is no clear dispersion relation, see Fig.~\ref{fig:fig6}(a).
Thus, from the single-particle   spectrum, it is hard to extract useful information on the kinetics of the doublons and holons.
On the other hand, when the HHG spectrum is measured, one finds almost the same result between the cases with the initial state determined from $\hH_{\rm spin}=\hH_{\rm spin,ex}+h_z \sum_i (-)^i \hs_{z,i} $ and the antiferromagnetic initial state, see Figs.~\ref{fig:fig6}(b) and (c).
This result is consistent with the fact that in $\hH_{\rm Mott,1}$, the charge dynamics concerning with the doublons and holons  is independent of the spin configuration.\cite{Koch1997}
The above results demonstrate that the HHG spectrum directly reflects the doublon-holon dynamics, but the single-particle relation does not necessarily so.
Thus, the relation between the HHG spectrum and the single-particle spectrum can be 
very different from what we expect in a semiconductor system.~\cite{Murakami2018PRL}
Our findings indicate that the HHG spectrum is the more direct tool to study the kinetics of the doublons and holons 
in a Mott insulator. 

 %%%%%%%%%%%%%%%%%%%%%%%%%%%%%%%%%%%%%%%%%%%%%
%%%%%%%%%%%%%%%%%%%%%%%%%%%%%%%%%%%%%%%%%%
A lesson from our HHG analysis of the 1d Mott insulator is that the three step model \eqref{eq:semi_classic} can be useful even for strongly correlated systems if we use a proper 
dispersion relation $\epsilon_g(p)$ related to many-body elementary excitations.
Here, the dispersion $\epsilon_g(p-A)$ describes the change of the energy under an adiabatic change of $A$, for the doublon-holon pair with the half relative momentum $p$ at $A=0$.
It is an interesting question if the same idea applies to other types of SCESs. Namely, one may obtain  $\epsilon_g(-A)$ by following the change of the energy of an excited state under an adiabatic change of $A$ and use it within the three step model to explain the HHG features.

\section{Conclusions}\label{sec:conclude}

In this paper, we studied HHG in the 1d Mott insulator described by the single-band Hubbard model using iTEBD and exact diagonalization.
We pointed out that the HHG originates from the doublon-holon recombination, at least when the gap is large enough, and demonstrated that the subcycle features are reasonably well captured by the semiclassical three step analysis for doublon-holon pairs.
The dynamics of a doublon-holon pair is ruled by the dispersion of the doublon-holon pair with respect to its relative momentum, which is not necessarily captured by the
single-particle  spectrum due to the many-body nature of the elementary excitations. Our results indicate that HHG in Mott insulators can be applied for a spectroscopy to directly measure the dispersion of the relevant many-body elementary excitations, here the doublon-holon pairs. 

Moreover, we introduced effective models based on the Schrieffer-Wolff transformation, which allows us to identify processes similar to and different from the semiconductor models,  and to discuss the role of these individual processes.
We showed that the spin dynamics, which has no analogue in semiconductor systems, substantially reduces the HHG intensity. 
This result indicates that the HHG intensity in Mott systems can be sensitive to spin dynamics. It will be interesting to study the effects of different types of spin couplings and  to discuss the controllability of the HHG spectrum via external magnetic fields.
Furthermore, we revealed the importance of the long-range component of the ``dipole moment" between the doublon band and the holon band for the HHG intensity, as well as the role of the correlated hopping of doublons and holons for the shape of the HHG spectrum.  
We expect that our results are also relevant for charge transfer (CT) insulators, although a detailed study of HHG in these systems is required for a precise statement.
 Candidates of 1d Mott insulators and CT insulators range from organic crystals, e.g., $\rm{ET}$-$\rm{F_2TCNQ}$, to cuprates, e.g., $\rm{Sr_2CuO_3}$.\cite{Oka2012PRB}
 It would be interesting to experimentally explore the HHG in these systems, and to compare the measurements with our theoretical predictions. 
%Candidate 1d Mott insulators range from organic crystals to cuprates such as $\rm{ET}$-$\rm{F_2TCNQ}$ and $\rm{Sr_2CuO_3}$,\cite{Oka2012PRB} and our results should be relevant for HHG in these systems.

For understanding the detailed relation between the elementary excitations and HHG in various SCESs, it is an interesting future problem to introduce concepts similar to doublon-holon pairs with different relative
momenta, i.e. a series of states that is connected via adiabatic changes of the vector potential and their dispersions, to other SCESs systems such as dimer-Mott
insulators.\cite{Ishihara2020}
This would help us to explore the spectroscopic application of HHG to detect the dynamics of these elementary excitations. 
Also, in many-body systems, elementary excitations can strongly interact with each other, as in the doublon-holon dynamics in a correlated spin background\cite{Denis2014PRB} or the singlon-triplon string state in multi-orbital systems.\cite{Markus2020}
Such effects may be taken into account as an extra potential between elementary excitations; for example, in the spin background, the potential should be proportional to the distance between the doublon and holon.
How this affects the dynamics of elementary excitations under strong fields and consequently the HHG spectrum is an interesting open question.
In addition, it is also important to understand the behavior of HHG in SCESs in a wider range of excitation frequencies.

Last but not least, in this study, we have introduced a new numerical method to measure the fermionic single-particle  spectrum within iTEBD. 
The method has advantages in that i) a momentum- and energy-resolved spectrum is obtained from a single-shot calculation and ii) it can be directly evaluated for a system in the thermodynamic limit.
The idea can be extended to nonequilibrium situations, and we expect future applications to various nonequilibrium evolutions of the single-particle  spectrum.

\acknowledgements
We would like to acknowledge useful discussions with Takashi Oka and Zala Lenar\v{c}i\v{c}.
This work was supported by a Grant-in-Aid for Scientific Research from JSPS, KAKENHI Grant Nos. JP19K23425, JP20K14412, JP20H05265 (Y.M.), JP19H05821, JP18K04678, JP17K05536 (A.K.), JST CREST Grant No. JPMJCR1901 (Y.M.), JPMJCR19T3 (S.T.), and
ERC Consolidator Grant No. 724103 (P.W.).
Some of the numerical calculations were performed on the Beo05 cluster at the University of Fribourg. 

\appendix

\section{Evaluation of the single-particle spectrum in iTEBD}\label{sec:iTEBD_Akw}
Here, we explain how to measure the single-particle Green's function for fermions,
\eqq{
G^R_{ij}(t) = -i\theta(t)\langle [\hc_i(t), \hc^\dagger_j(0)]_+  \rangle,
}
within iTEBD.
Using DMRG or ED for finite systems, one can directly apply an operator 
$\hc^\dagger_j(0)$ to the wave-function and measure $\hc_i(t)$ to 
evaluate the single-particle Green's function of finite size systems in principle. 
In the case of iTEBD, using the infinite boundary condition,\cite{Phien2012PRB} 
we can calculate the spectral functions 
in spin systems~\cite{Takayoshi2018PRB,Suzuki2018PRB} 
without finite size effects, which corresponds to applying a bosonic operator to the system.
However, it is not straightforward to apply a fermionic operator to the 
matrix product state while keeping its canonical form.
In the following, we describe how to measure the single-particle 
Green's function via a pump-probe simulation by considering an auxiliary band and measuring a nonlocal correlation function of fermionic operators. 

In general, the pump-probe approach allows us to directly measure the linear response function,
\eqq{
\chi^R_{BA}(t) = -i  \theta(t) \langle  [\hB(t),\hA(0)]\rangle_{\rm sys}.
}
Here, we express the unperturbed Hamiltonian for the system of interest as $\hH_{\rm sys}$,  $\langle \rangle_{\rm sys}$ denotes the expectation value for an equilibrium state of $\hH_{\rm sys}$, and $\hA$ and $\hB$ are some operators. 
$\chi^R_{BA}(t)$ dictates the change of the expectation value of $\hB$ ($\delta B(t)$) induced by the small perturbation from $\hH_{\rm ex}=\delta F(t) \hA$:
\eqq{
\delta B(t)= \int d\bar{t} \chi^R_{BA}(t-\bar{t})\delta F(\bar{t}).
}
The Fourier components satisfy $\delta B(\omega) = \chi^R_{BA}(\omega)\delta F(\omega)$.
One can evaluate $\chi^R_{BA}(t)$ directly by the real time evolution of a weak-enough excitation.
This can be utilized in any type of real-time numerical simulation to evaluate the response functions, see e.g., Refs. ~\onlinecite{Murakami2016,Werner2019} for applications in the context of nonequilibrium Green's function methods.

On the other hand, with iTEBD, the wave function $|\Psi(t)\rangle$ is expressed as a matrix product state in the canonical form. Using this fact one can efficiently evaluate equal-time nonlocal correlation functions,
\eqq{
X_{\beta\alpha,ij}(t) \equiv \langle \Psi(t) |\hat{\beta}_i \hat{\alpha}_j |\Psi(t)\rangle,
}
regardless of the type of $\hat{\beta}_i$ and $\hat{\alpha}_j$, i.e., fermionic or bosonic.

We use the above two points to measure $G^R_{ij}(t)$.
First, we separate $G^R_{ij}(t)$ into two parts as $G^R_{ij}=G^{R1}_{ij}+ G^{R2}_{ij}$, where $G^{R1}_{ij}(t) =  -i\theta(t)\langle \hc_i(t) \hc^\dagger_j(0)  \rangle$ and $G^{R2}_{ij}(t) =  -i\theta(t)\langle  \hc^\dagger_j(0) \hc_i(t)  \rangle$.
Then, we consider an auxiliary band of fermions $\hH_{\rm aux}=\omega_0\sum_i \hb_i^\dagger \hb_i$. 
The total Hamiltonian without external perturbation is $\hH_{\rm tot} = \hH_{\rm sys} + \hH_{\rm aux}$. 
We take the initial state $|\Psi_{\rm tot}\rangle$ as 
\eqq{
|\Psi_{\rm tot}\rangle = |\Psi_{\rm sys}\rangle \otimes |{\rm vac}\rangle_{\rm aux},
}
where $|{\rm vac}\rangle_{\rm aux}$ is the vacuum state of the auxiliary band and $|\Psi_{\rm sys}\rangle$ is the ground state of the system.
Next, we weakly excite the system by applying a homogeneous field $\delta F(t)\sum_l \hat{A}_l $with $\hat{A}_l = \hb^\dagger_l 
\hc_l+\hc^\dagger_l\hb_l$ at time $0$ and observe $\hat{B}_{ji}=\hb_j^\dagger \hc_i$ at time $t$.
This procedure enables us to measure 
\eqq{
\tilde{\chi}_{ij}(t) & \equiv -i\theta(t) \Big\langle \Big[\hat{B}_{ji}(t), \sum_l \hat{A}_l (0)\Big]\Big\rangle_{\rm tot} \nonumber\\
&= i\theta(t)e^{i\omega_0 t}\langle \hc^\dagger_j(0)\hc_i(t) \rangle_{\rm sys} = -e^{i\omega_0 t} G_{ij}^{R2}(t),
}
and we obtain $G_{ij}^{R2}(t)$ by setting $\omega_0=0$.
Note that since the canonical form of the matrix product for $|\Psi_{\rm 
sys}\rangle \otimes |{\rm vac}\rangle_{\rm aux}$ is easily expressed by 
taking the direct product of the canonical representation for 
$|\Psi_{\rm sys}\rangle$ and $|{\rm vac}\rangle_{\rm aux}$,
we do not need to deal with $\hH_{\rm tot}$. This fact saves the computational cost for the preparation of the initial state.

In the same way, we can calculate $G_{ij}^{R1}(t)$ by exciting the 
system with  $\delta F(t)\sum_l \hat{A}'_l $ and  $\hat{A'}_l = \hb_l \hc_l+\hc^\dagger_l\hb^\dagger_l$ at 
time $0$ and observing $\hat{B}_{ji}=\hb_j  \hc_i$ at time $t$.
This process can be circumvented in special cases where $G_{ij}^{R1}(t)$ can be related to  $G_{ij}^{R2}(t)$.
For example, the half-filled Hubbard model is symmetric under $\hc^\dagger_{i\sigma} \rightarrow (-)^i \hc_{i\sigma}$. 
If this symmetry is not broken in the ground state, we have 
\eqq{
G^{R1}_{ij}(t) = -i\theta(t)   \langle  \hc_{i\sigma}(t)  \hc^\dagger_{j\sigma}(0)\rangle  = -(-)^{j-i} {G_{ij}^{R2}}^{*}(t).
}

Although precise single-particle spectra in equilibrium can be obtained by dynamical DMRG,\cite{White1999} its application is limited to finite systems.
Our method has the advantage that it is directly applicable to the thermodynamic limit and that it can be easily extended to nonequilibrium situations.

%%%%%%%%%%%%%%%%%%%%%%%%%%%%%%%%%%%%%%%%%%%%%%%%%%%%%%%%%%%%%
%%%%%%%%%%%%%%%%%%%%%%%%%%%%%%%%%%%%%%%%%%%%%%%%%%%%%%%%%%%%%
\section{Higher order correction of the effective model}\label{sec:Appendix_eff}
The higher order terms of $\hS$ and $\hH_{\rm Mott}$ can be obtained %deductively 
recursively 
using the following fact:\cite{MacDonald1988}
If an operator $\hat{M}_n$ changes the number of doublons by $n$ and does not change the total number of charges, 
we have 
\eqq{
[\hH_U, \hat{M}_n] = nU \hat{M}_n.
}

The procedure for determining $\hS^{(i)}$ and $\hH_{\rm Mott,i}$ is the following:
\begin{enumerate}
\item Write down the components in $\hH_{\rm Mott}$ of the order $\mathcal{O}(U(\frac{v}{U})^i)$, which involve $i[\hS^{(i)}(t),\hH_U]$.  We denote the components except for $i[\hS^{(i)}(t),\hH_U]$ by $\hat{M}^{(i)}(t)$.
\item We use $\hat{M}^{(i)}_n(t)$ to denote the terms in $\hat{M}^{(i)}(t)$ that change the number of doublons by $n$.
\item $\hS^{(i)}$ is given by 
\eqq{
\hS^{(i)}(t) = \frac{-i}{U}\sum_{n\neq0} \frac{1}{n} \hat{M}^{(i)}_n(t) \label{eq:S_i}.
}
\item  $\hH_{\rm Mott,i}(t) = \hH_{\rm Mott,i-1}(t) + \hat{M}^{(i)}_0(t)$.
\end{enumerate}
One can directly see that with Eq.~\eqref{eq:S_i}, $i[\hS^{(i)},\hH_U]$  cancels with $\hat{M}^{(i)}_n$ for $n\neq 0$.

For example, the component of $\hH_{\rm Mott}$ for $\mathcal{O}(U(\frac{v}{U})^2)$ can be expressed as 
\eqq{
& i[\hS^{(1)}(t),\hH_{\rm kin,0}(t)] + \frac{i}{2} [\hS^{(1)}(t),\hH_{\rm kin,\pm}(t)] + i[\hS^{(2)}(t),\hH_U]
}
with $\hH_{\rm kin,0}\equiv \hH_{\rm kin,LHB} + \hH_{\rm kin,UHB}$ and $\hH_{\rm kin,\pm}=\hH_{\rm kin,+} + \hH_{\rm kin,-}$.
Then, we have
\eqq{
\hM^{(2)}_1(t) &= \frac{1}{U}[\hH_{\rm kin, +}(t), \hH_{\rm kin,0}(t)] , \nonumber\\
\hM^{(2)}_0(t) &= \frac{1}{U} [\hH_{\rm kin,+}(t),\hH_{\rm kin,-}(t)], \\
\hM^{(2)}_{-1}(t) & =\hM^{(2)}_1(t)^\dagger .\nonumber 
}

More explicitly, we have 
   \eqq{
&   \hM^{(2)}_1(t) = \nonumber\\ 
&  -2\frac{v^2}{U} \sum_{i,\sigma} [ \hh^\dagger_{i+1,\sigma} \hc^\dagger_{i,\sigma} \hc^\dagger_{i,\bar{\sigma}} \hh^\dagger_{i-1,\bar{\sigma}}
+  \hd^\dagger_{i+1,\bar{\sigma}} \hc_{i,\sigma} \hc_{i,\bar{\sigma}}  \hd^\dagger_{i-1,\sigma} ] \nonumber \\
&  +  \frac{v^2}{U} \sum_{i,\sigma} \sum_{m=\pm1} e^{2iA\cdot m}  \big[ 2  \hd^\dagger_{i+m,\bar{\sigma}} \hc^\dagger_{i,\bar{\sigma}} c_{i,\sigma} \hh^\dagger_{i-m,\bar{\sigma}} \\
&  \;\;\;\;\;\;\;\;\;\;\;\; \;\;\;\; \;\;\;\;\;\;\;\;\;\;\;\; \;\;\;\; +  (\bar{n}_{i,\bar{\sigma}}-n_{i,\bar{\sigma}}) \hd^\dagger_{i+m,\bar{\sigma}}  \hh^\dagger_{i-m,\sigma}\bigl] .\nonumber  }
 Also, we have 

 \eqq{
  \hM^{(2)}_0(t) &= \hH^{(2)}_{\rm kin,LHB} +  \hH^{(2)}_{\rm kin,UHB} + \hH^{(2)}_{\rm U,shift} \nonumber\\
       &\;\;\;+  \hH_{\rm spin,ex} +  \hH_{\rm dh,ex} +  \hH^{(2)}_{\rm dh,slide},
 }
 with 
 
  \eqq{
  &\hH^{(2)}_{\rm kin,LHB}  =  -\frac{J_{\rm ex}}{4}\sum_{i,\sigma} \sum_{m=\pm 1} e^{2iA\cdot m} \nonumber\\
 & \;\;\; \Bigl[ n_{i,\bar{\sigma}} \hh_{i+m,\sigma}  \hh^\dagger_{i-m,\sigma} - \hc^\dagger_{i,\sigma}\hc_{i,\bar{\sigma}} \hh_{i+m,\bar{\sigma}}  \hh^\dagger_{i-m,\sigma}  \Bigl],
  }
 
 \eqq{
&  \hH^{(2)}_{\rm kin,UHB}  = \frac{J_{\rm ex}}{4}\sum_{i,\sigma} \sum_m e^{2iA\cdot m} \nonumber\\
 &  \Bigl[\bar{n}_{i,\bar{\sigma}}  \hd^\dagger_{i+m,\bar{\sigma}} \hd_{i-m,\bar{\sigma}}+ \hc^\dagger_{i,\bar{\sigma}}\hc_{i,\sigma}  \hd^\dagger_{i+m,\bar{\sigma}} \hd_{i-m,\sigma}  \Bigl],
 }

 \eqq{
  \hH^{(2)}_{\rm U,shift} & =J_{\rm ex} \sum_i (\hn_{i\uparrow}-\frac{1}{2})(\hn_{i\downarrow}-\frac{1}{2}),
  }
  \eqq{
  \hH_{\rm dh,ex} & = -\frac{J_{\rm ex}}{2}\sum_i [e^{-2iA} \heta^+_i \heta^-_{i+1} + e^{2iA}  \heta^-_i \heta^+_{i+1} + 2\heta^z_i \heta^z_{i+1}],
  }
 
 \eqq{
  \hH^{(2)}_{\rm dh,slide} & = -\frac{J_{\rm ex}}{4}\sum_{i,\sigma}\sum_{m=\pm1}  \Bigl [ \hh^\dagger_{i+m,\sigma}  \hc_{i\sigma}^\dagger \hc^\dagger_{i,\bar{\sigma}}  \hd_{i-m,\sigma} + h.c.].
 }
 
 Here, we introduced the $\eta$-operators\cite{Yang1989PRL,Li2020arxiv} as $\heta^+_i = \theta_i \hc^\dagger_{i\uparrow}\hc^\dagger_{i\downarrow}$, $\heta^-_i = \theta_i \hc_{i\downarrow}\hc_{i\uparrow}$ and $\heta^z_i =\frac{1}{2}(\hn_i-1)$, and $\hH_{\rm spin,ex} $ is given in Eq.~\eqref{eq:H_spin}.
The meaning of each term is as follows. $\hH^{(2)}_{\rm kin,LHB}$ describes the hopping of a holon and  $\hH^{(2)}_{\rm kin,UHB}$ describes the hopping of a doublon. $\hH^{(2)}_{\rm U,shift}$ describes the shift of the local interaction $U$, while $\hH_{\rm dh,ex}$ is the exchange coupling of the doublon and holon. $\hH^{(2)}_{\rm dh,slide}$ describes the simultaneous hopping of a doublon and a holon to the neighboring sites.

 For the case of $i=3$, we have
\eqq{
\hM^{(3)}_{+2}(t) &=  \frac{1}{2U}[ M_1 ^{(2)} (t), H_{\rm kin,+} (t)] \\
\hM^{(3)}_{+1}(t) 
 &= \frac{2}{3U}[\hH_{\rm kin,+}, M_0 ^{(2)} (t)]  + \frac{1}{U}[ M_1 ^{(2)} (t), H_{\rm kin,0}] \nonumber \\
\hM^{(3)}_{0}(t) 
&=  \frac{1}{2U} [\hM_1 ^{(2)} (t),   H_{\rm kin,-} (t)]  - \frac{1}{2U} [\hM_{-1} ^{(2)} (t),   H_{\rm kin,+} (t)] \nonumber 
}
and $\hM^{(3)}_{-1}(t) =\hM^{(3)}_{1}(t) ^\dagger$ and $\hM^{(3)}_{-2}(t) =\hM^{(3)}_{2}(t) ^\dagger$.

As for $\hH_{\rm ex}$, we can expand it as 
 \eqq{
\hH_{\rm ex}(t) = -\dot{\hS} -\frac{i}{2} [\hS,\dot{\hS}] + \frac{1}{6} [\dot{\hS}\hS\hS -2 \hS\dot{\hS}\hS + \hS\hS\dot{\hS}] + \cdots,
 }
 where $\dot{\hS} = \partial_t \hS(t)$. 
 Thus, we have 
  \eqq{ -\dot{\hS}^{(1)}}
 for  $O(E_0\cdot \frac{v}{U})$,
 \eqq{ -\dot{\hS}^{(2)}-\frac{i}{2} [\hS^{(1)},\dot{\hS}^{(1)}] }
 for $O(E_0\cdot (\frac{v}{U})^2)$
 and 
\eqq{ &-\dot{\hS}^{(3)} -\frac{i}{2} [\hS^{(1)},\dot{\hS}^{(2)}] -\frac{i}{2} [\hS^{(2)},\dot{\hS}^{(1)}] \\
 &+ \frac{1}{6} [\dot{\hS}^{(1)}\hS^{(1)}\hS^{(1)} -2 \hS^{(1)}\dot{\hS}^{(1)}\hS^{(1)} + \hS^{(1)}\hS^{(1)}\dot{\hS}^{(1)}]  \nonumber
 }
for $O(E_0\cdot (\frac{v}{U})^3)$. Note that $ -\dot{\hS}^{(1)}$ includes doublon-holon pair creation at the nearest neighbor sites.
As one can see from the form of $\hM^{(2)}_1$, $ -\dot{\hS}^{(2)}$ includes  doublon-holon pair creation at the next-nearest neighbor sites.
Similarly, $-\dot{\hS}^{(3)}$ includes  doublon-holon pair creation at the 3rd-nearest neighbor sites.

%%%%%%%%%%%%%%%%%%%%%%%%%%%%%%%%%%%%%%%%%%%%%
 \begin{figure}[t]
  \centering
    \hspace{-0.cm}	
    \vspace{0.0cm}
  \includegraphics[width=85mm]{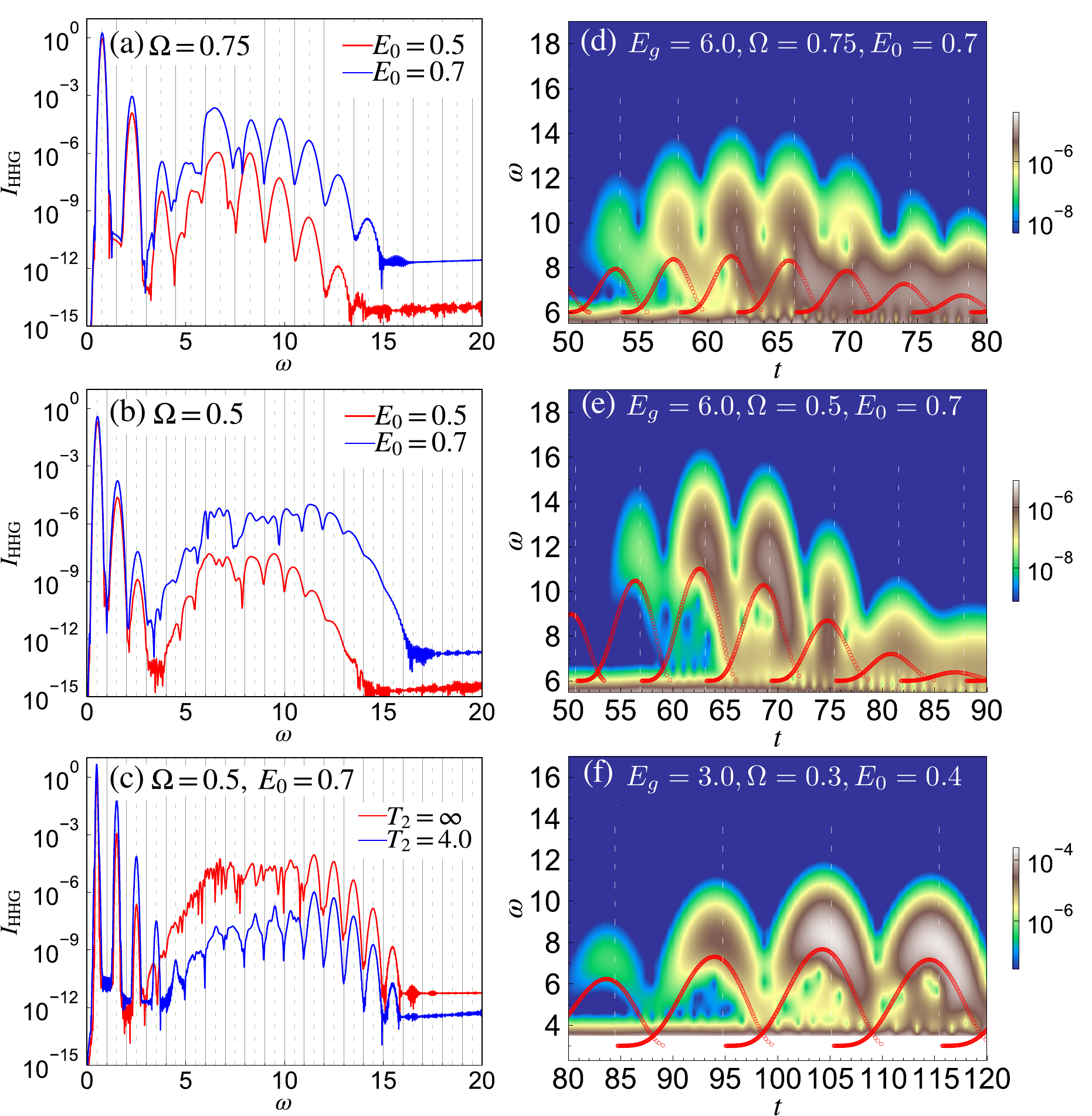} 
  \caption{(a) (b) HHG spectra of the semiconductor with $E_g=6.0$ for (a) $\Omega=0.75$ and (b) $\Omega=0.5$ for the specified field strengths. Here, we use $t_0=60$ and $\sigma=15$ for the pulse and  $\sigma'=20$ for the Fourier transformation (FT).
  (c) HHG spectra of the semiconductor with $E_g=6.0$ for $\Omega=0.5, E_0=0.7, t_0=160,\sigma=40$ with and without the phenomenological dephasing term ($T_2$). Here, $\sigma'=60$ is used for the FT.
  (d)(e) Subcycle analysis of the HHG signal. $I_{\rm HHG}(\omega,t_p)$ is plotted for  (d) $\Omega=0.75$ and (e) $\Omega=0.5$. Here, we use $E_g=6.0$, $E_0=0.7$, $t_0=60$ and $\sigma=15$ for both cases. (f) %Subcycle analysis of the HHG, 
  $I_{\rm HHG}(\omega,t_p)$ for $E_g=3.0$. The parameters of the pump are $\Omega=0.3,E_0=0.4,t_0=100$ and $\sigma=30$. For the subcycle analysis $\sigma'=0.8$ is used. The red markers indicate the results of the subcycle analysis ($\omega_{\rm emit}(t_{\rm rec})$) with the dispersion $\epsilon_{a}(p)=\pm \frac{D}{2} \pm 2v_a \cos(p)$. The results in  panels (a)(b)(d)(e)(f) are obtained without phenomenological dephasing term ($T_2=\infty$).}
  \label{fig:fig_A1}
\end{figure}
%%%%%%%%%%%%%%%%%%%%%%%%%%%%%%%%%%%%%%%%%%

\section{Semiconductor models}\label{sec:semicon}
In this section, we show how the HHG results of the simplest but typical semiconductor model look like in the parameter regime relevant for this paper.
As the semiconductor model, we consider the spin-less two-band model in the dipole gauge,\cite{Higuchi2014,Li2020,Murakami2020}  
\begin{align}
\hat{H}_{\rm semi}&=- \sum_{i ,a=c,v}  [v_a e^{-iA(t)} \hc^\dagger_{i,a} \hc_{i+1,a} + h.c.]  \label{eq:semiconductor} \\
&+\frac{D}{2}\sum_{i} [\hn_{i,c} - \hn_{i,v}] -E(t) \sum_{i,a} d_{a} \hc^\dagger_{i,a} \hc_{i,\bar{a}}. \nonumber
\end{align}
Here, $a$ is the index for the conduction band ($c$) or the valence band ($v$), $D$ is the band level difference and $d_a$ is the dipole moment,
which is assumed to be local.
The light-matter coupling is taken into account by the Peierls phase and the dipole excitation term (the last term). 
The size of the latter determines the transitions between the conduction band and the valence band, and the size of the dipole moment can be usually comparable but no more than the lattice constant. 
Here, we assume that it is the same as the lattice constant, $d_a=1$.
We study this model by solving the equation of motion for the single-particle density matrix, i.e. the semiconductor Bloch equation.\cite{Vampa2014PRL,Vampa2015PRB,Murakami2020}
Note that the equations of motion in this gauge are essentially identical to those for the Hamiltonian in the length gauge,\cite{Vampa2014PRL,Vampa2015PRB} 
so that the semiclassical theory derived in Ref.~\onlinecite{Vampa2015PRB} is directly applicable.
Additionally, in the semiconductor Bloch equation, one can phenomenologically take account of the dephasing effects originating from electron-phonon coupling, disorder and electron-electron interactions via 
the relaxation time approximation, which involves  the dephasing time $T_2$.\cite{Vampa2014PRL,Vampa2015PRB} ($T_2=\infty$ means no dephasing).
In the following, we set $v_c=1$ and $v_v=-1$, where the system has a direct band gap.
We note that, conceptually, the band level difference $D$ corresponds to the local Coulomb interaction $U$ in the Hubbard model.
The band gap $E_g$ is $D-2(v_c-v_v)$. 

In Figs.~\ref{fig:fig_A1}(a,b,d,e), we show the HHG spectrum, which corresponds to Fig.~\ref{fig:fig2}(a)(b), and the results of the subcycle analysis, which correspond to Fig.~\ref{fig:fig3}(c)(d).
For $\Omega=0.75$, one can clearly see the HHG signals at odd frequencies [Fig.~\ref{fig:fig_A1}(a)].
On the other hand, for $\Omega=0.5$ the HHG peaks are less clear and  some of them deviate from the odd harmonics [Fig.~\ref{fig:fig_A1}(b)].
This can be attributed to the short pulse and the lack of dephasing in the present model \eqref{eq:semiconductor}, i.e. $T_2=\infty$. 
Indeed, with a longer pulse, the peaks become sharper around the upper edge of the plateau  [Fig.~\ref{fig:fig_A1}(c)], 
while the peaks around the gap edge become ill-defined without the dephasing.
With the dephasing, the latter peaks become sharper.\cite{Vampa2014PRL,Vampa2015PRB} 
As for the subcycle analysis, some subcycle features are similar to those in the Mott insulators, see Fig.~\ref{fig:fig_A1}(d)(e).
The semiclassical model captures several features such as the timing of the emission of high frequency light, and the level of agreement with the actual results is similar to the case of the Mott insulator.
The agreement with the semiclassical theory tends to become better as we decrease the gap and the excitation frequency, see for example Fig.~\ref{fig:fig_A1}(f).

\bibliography{HHG_Ref}

\end{document}